\documentclass[lettersize,journal]{IEEEtran}
\usepackage{algorithmic}
\usepackage{algorithm}
\usepackage{array}
\usepackage[caption=false,font=normalsize,labelfont=sf,textfont=sf]{subfig}
\usepackage{textcomp}
\usepackage{stfloats}
\usepackage{url}
\usepackage{verbatim}
\usepackage{graphicx}
\usepackage{cite}

\usepackage{amsmath,amssymb,amsfonts}
\usepackage{textcomp}
\usepackage{color}
\usepackage{mathtools}
\usepackage{dsfont} 

\usepackage{array}
\newcolumntype{L}{>{\centering\arraybackslash}m{1.5cm}}
\newcolumntype{M}{>{\centering\arraybackslash}m{1.5cm}}
\newcolumntype{N}{>{\centering\arraybackslash}m{2.3cm}}

\usepackage[skip=2pt,font=small]{caption}

\newtheorem{theorem}{Theorem}[section]

\newtheorem{lemma}[theorem]{Lemma}

\usepackage{cite}
\newcommand{\muoneone}{\mu_{11}}
\newcommand{\muonetwo}{\mu_{12}}
\newcommand{\mutwoone}{\mu_{21}}
\newcommand{\mutwotwo}{\mu_{22}}

\allowdisplaybreaks

\newcommand{\pr}{\mathbb{P}}

\begin{document}
	\title{On the Interplay of Clustering and Evolution in the Emergence of Epidemic Outbreaks
}
	\author{\IEEEauthorblockN{Mansi Sood$^\gamma$, Hejin Gu$^\gamma$, Rashad Eletreby$^\dagger$, Swarun Kumar$^\gamma$, Chai Wah Wu$^{*}$, Osman Ya\u{g}an$^\gamma$}\\
		\IEEEauthorblockA{$^\gamma$Department
			of Electrical and Computer Engineering, \\
			Carnegie Mellon University, Pittsburgh,
			PA 15213 USA\\
   $^\dagger$Walmart Global Tech. 221 River St, Hoboken, NJ 07030 USA\\
			$^*$IBM Research, Thomas J. Watson Research Center, Yorktown Heights, NY 10598 USA\\
			{{msood@andrew.cmu.edu, hejing@alumni.cmu.edu, eletreby.rashad@gmail.com, swarun@cmu.edu, cwwu@us.ibm.com, oyagan@andrew.cmu.edu}}}}
	\maketitle
	\begin{abstract}
In an increasingly interconnected world, a key scientific challenge is to examine mechanisms that lead to the widespread propagation of contagions, such as misinformation and pathogens, and identify risk factors that can trigger large-scale outbreaks. Underlying both the spread of disease and misinformation epidemics is the \emph{evolution} of the contagion as it propagates, leading to the emergence of different \emph{strains}, e.g., through genetic mutations in pathogens and alterations in the information content. Recent studies have revealed that models that do not account for heterogeneity in transmission risks associated with different strains of the circulating contagion can lead to inaccurate predictions. However, existing results on \emph{multi-strain} spreading assume that the network has a vanishingly small \emph{clustering} coefficient, whereas clustering is widely known to be a fundamental property of real-world social networks. In this work, we investigate spreading processes that entail evolutionary adaptations on random graphs with tunable clustering and arbitrary degree distributions. We derive a mathematical framework to quantify the epidemic characteristics of a contagion that evolves as it spreads, with the structure of the underlying network as given via arbitrary {\em joint} degree distributions of \emph{single-edges} and \emph{triangles}. To the best of our knowledge, our work is the first to jointly analyze the impact of clustering and evolution on the emergence of epidemic outbreaks. We supplement our theoretical finding with numerical simulations and case studies, shedding light on the impact of clustering on contagion spread.
	\end{abstract}
	\section{Introduction}
\subsection{Background and Related Work}
The recent COVID-19 outbreak, caused by the novel coronavirus SARS-CoV-2, resulted in a catastrophic loss of life and  disrupted economies around the world \cite{msemburi2023estimates}. The highly  transmissible and rapidly \emph{evolving} \cite{harvey2021sars} nature of the SARS-CoV-2 coronavirus led to an unprecedented burden on critical healthcare infrastructure. Moreover, the absence of pharmacological interventions in the early stages of the outbreak led to the adoption of \emph{nonpharmaceutical interventions} that alter the contact patterns to combat the spread. Understanding the impact of countermeasures that alter the patterns of interaction requires network-based approaches for understanding spreading phenomena. In addition to the high rates of transmission of infection, the spread of misinformation pertaining to COVID-19 on social media platforms further impeded efforts to mitigate its spread \cite{papakyriakopoulos2020spread}. An investigation into the mechanisms underlying the spread of contagions across these diverse contexts such as the spread of infectious diseases and misinformation, has received broad attention under the frameworks of network-based {epidemiological models}, {spreading} processes and {cascades} \cite{pappas-survey,rho-survey,miller-textbook,ne:newman2018networks}.


		\begin{figure}[t]
		\centering
		\includegraphics[scale=0.23]{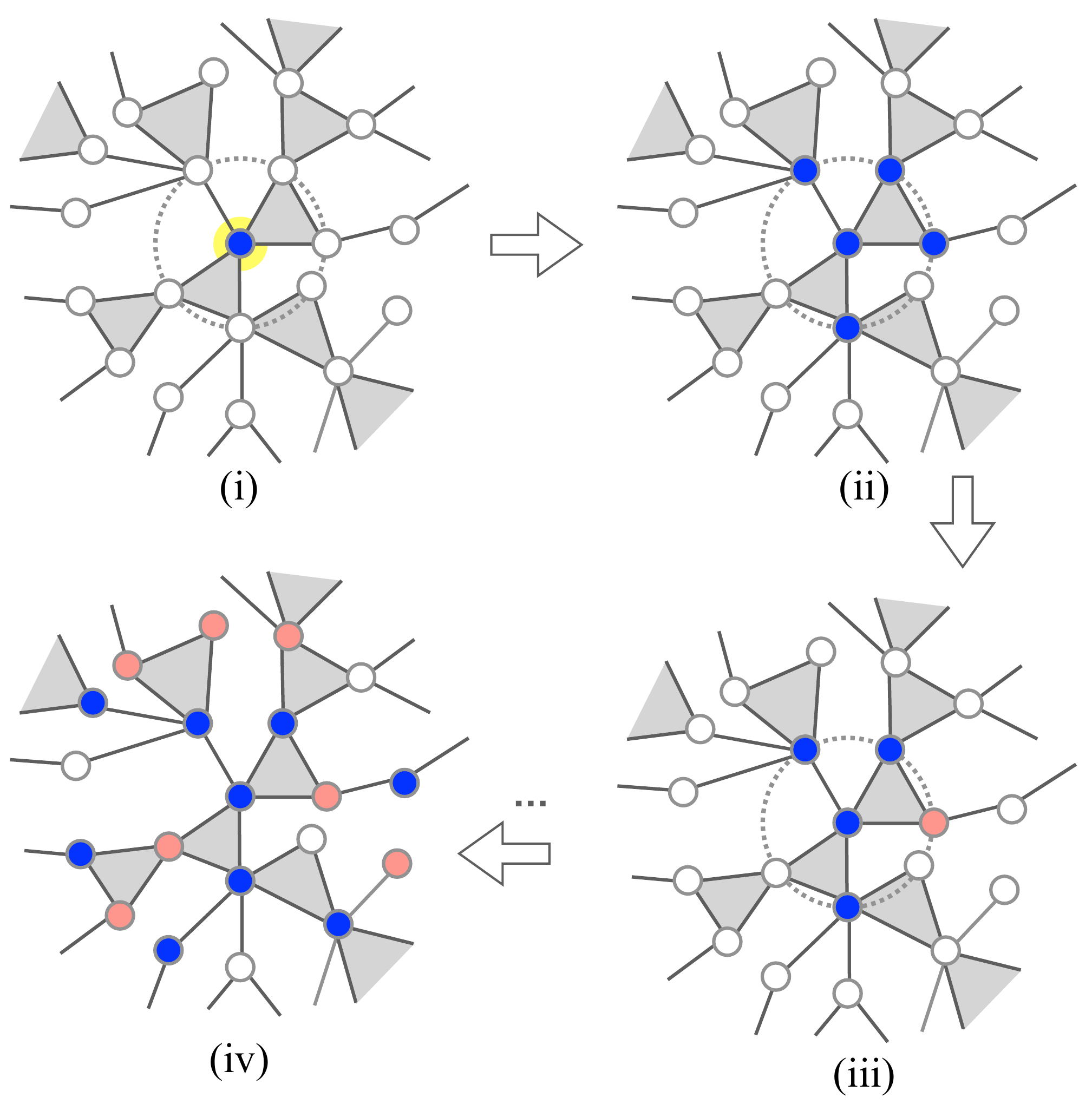}
		\caption{\sl An illustration of the multi-strain spreading on a clustered random network. The contact network comprises a clustered network where the number of single-edges and triangles attached to each node is separately specified through the joint degree distribution for the configuration model. We consider the propagation of two strains (strain-$i$, $i=1,2$) of a contagion indicated in blue and red. The spreading proceeds as follows: i) Initially all nodes are in the susceptible state and an arbitrarily chosen seed node acquires strain-$i$. ii) The seed node independently infects its susceptible neighbors with a probability corresponding to the strain it carries ($T_i$). iii) After transmission, the contagion mutates to strain-$j$ within the newly infected hosts independently with probability $\mu_{ij}$. iv) The process continues recursively and terminates when no further infections are possible.}
		\label{fig:0}
		\vspace{-4mm}
	\end{figure}
 A key determinant of the speed and scale of propagation of a contagion is its propensity to get altered upon interactions with different hosts {\cite{girvan2002simple,antia2003role,multistrain-review, marquioni2021modeling, ne:alexday,ne:pnas, zhang2021epidemic,Xue2024pnasref,rudiger2020epidemics,leventhal2015evolution,buckee2004effects}}. In the context of transmission of infectious diseases, as pathogens enter new species, they undergo evolutionary adaptations or \emph{mutations} to better adapt to the new host environment {\cite{antia2003role,zhang2021epidemic,ne:alexday,marquioni2021modeling,buckee2004effects,girvan2002simple}}. The different {\em variants} or {\em strains} arising from {mutations} of the pathogen have varying risks of transmission, commonly measured through the reproduction number or $R_0$, which quantifies the mean number of secondary infections triggered by an infected individual \cite{AndersonMay,Vespignanir0}. Moreover, even when a sizeable fraction of the population gains immunity through vaccination or natural infection, the emergence of new variants that can evade the acquired immunity poses a continued threat to public health \cite{islam2022new,tregoning2021variants}. Akin to different {strains} of a pathogen arising through evolutionary adaptations or mutations, different \emph{versions} or { variants} of the information are created as the content is altered on social media platforms \cite{Adamic2016,zhang2013rumor,ne:pnas}. For instance, in \cite{Adamic2016}, the authors characterize the mutation rate by using the \emph{edit distance} between a variant posted by a user and the variants previously posted by the user’s friends. The resulting {variants/strains} of the information may have varying propensities to be circulated in the social network \cite{ne:pnas}. A key scientific goal across the above spreading phenomena is to examine the underlying mechanisms that lead to the widespread propagation of contagions, e.g., misinformation and pathogens, and identify risk factors that can trigger widespread outbreaks. From the standpoint of spreading processes, this necessitates the development and analysis of a rich class of agent-based models that can simultaneously account for realistic patterns of interaction in the population and evolution in the contagion. 

Most commonly used network-based epidemiological models typically neglect the role of evolutionary adaptations in the spread of contagions \cite{newman-disease,kenah2007second,ne:newman2018networks,miller-textbook,miller2009percolation,NewmanRandomClustering}. However, as discussed above, in the context of the spread of infections (respectively, misinformation), different variants of the contagion emerge as it propagates in the network due to mutations in the pathogen (respectively, alteration in the information content). It is known \cite{buckee2004effects, ne:pnas,multilayermultistrain} that the host contact network structure shapes the emergence of pathogen strain structure and resulting dynamics. Moreover, recent studies \cite{ne:pnas, multilayermultistrain} have highlighted that models that do not consider evolutionary adaptations lead to incorrect predictions about the spreading dynamics of mutating contagions on contact networks. While a growing body of work {\cite{ne:alexday,ne:pnas, zhang2021epidemic,Xue2024pnasref,rudiger2020epidemics,buckee2004effects}} investigates the spread of evolving contagions on contact networks; however, these studies assume that the network admits a vanishingly small clustering coefficient in the limit of large network size. Hence, these models are limited in their expressibility and are unable to capture several important aspects of real-world social networks, most notably the property of high clustering \cite{serrano2006clustering}, which is known to have a significant impact on the behavior of spreading processes\cite{hackett2011cascades,huang2013robustness, Britton_clustering_2008, clustering-vaccination-math}. To better model social networks that are typically clustered, we employ a model of random networks with tuneable {\em clustering} and arbitrary degree distributions introduced by Miller \cite{miller2009percolation} and Newman \cite{NewmanRandomClustering}. We present a summary of key references in Table~\ref{table:tab1} and direct the reader to \cite{pappas-survey,rho-survey,miller-textbook,ne:newman2018networks} and additional references therein. 
			\begin{table}[htbp]
	\begin{center}
		\begin{tabular}{|L| L| L|N|} 
			\hline
			\textbf{Single-strain/ Multi-strain}\footnotemark & \textbf{Clustered/ not clustered\footnotemark } &   \textbf{Single-layer/ Multi-layer}\footnotemark&\textbf{Related Work}\footnotemark\\
			\hline\hline
			single-strain &  not clustered & single-layer & \cite{newman-disease,kenah2007second,ne:newman2018networks} \\ 
			\hline
			single-strain &  not clustered & {multi-layer} &  \cite{haccett-percolation-multiplex,Bianconi_2017, osman-yurun-2024,ne:conjoining}\\ 
			\hline
			{single-strain} &  {\emph{clustered}} & single-layer &  {\cite{miller2009percolation,NewmanRandomClustering,serrano2006clustering, clustering-vaccination-math, hackett2011cascades, volz-miller-2011effects, volz-miller-2011effects, mann2021percolation,huang2013robustness} }\\ 
			\hline
			{single-strain} &  {\emph{clustered}} & multi-layer & \cite{yong_tnse, mann-arbitrary-clustering, huang2013robustness} \\ 
			\hline
			\emph{multi-strain} & {not clustered} & single-layer & {\cite{ne:alexday,ne:pnas, zhang2021epidemic}} 
\\ 
			\hline
			\emph{multi-strain} &  {not clustered} & multi-layer & \cite{multilayermultistrain} \\ 
   \hline
   			\emph{multi-strain} &  {\emph{clustered}} & single-layer & our focus \\ 
			\hline
		\end{tabular}
	\end{center}
	\caption{\sl Overview of related works on network-based epidemiological models. Existing results for multi-strain spreading only focus on networks with vanishingly small clustering coefficients, while we account for multi-strain spreading in networks with tuneable clustering. }
	\label{table:tab1}
	\vspace{-3mm}
\end{table}
\footnotetext[2]{Note that we focus on the spread of contagions where a single infectious contact can lead to infection or awareness of a piece of information similar to the independent cascade models \cite{icltm-book-chapter}, but in contrast to models such as linear threshold models \cite{LTM} for influence maximization. }
\footnotetext[3]{We say that a network is \emph{not} clustered if the global clustering coefficient is vanishingly small in the limit of large network size.}
\footnotetext[4]{The single-layer model typically refers to a single network-layer generated using the {\em configuration} model. In contrast, the \emph{multi-layer} model \cite{multilayermultistrain, ne:conjoining} is typically constructed by taking the disjoint union $(\amalg)$ of network layers generated independently according to the {configuration} model.}
\footnotetext[5]{Note that we consider SIR epidemics where each node gets only one chance to infect their neighbors which is a critical assumption for the branching process formulation. A complementary line of work provides an exact analysis of continous time SIR processes modeled as high dimensional Markov chains, albeit more challenging to interpret given the complexity of the state space growing exponentially with the network size; we refer the reader to \cite{kavitha2023exact,pappas-survey} and the references therein for more details.}.
	\subsection{Main Contributions}
 To the best of our knowledge, our work is the first to analyze \emph{multi-strain spreading} on networks with \emph{tuneable clustering}. Our proposed framework (Section~\ref{sec:setup}) enables joint evaluation of the impact of evolutionary adaptations in the contagion and clustering in the contact network for arbitrary degree distributions and mutation patterns. With the aim of understanding the mechanisms underlying epidemics caused by contagions that get altered as they propagate, we derive key epidemiological quantities for the multi-strain model \cite{ne:alexday} on random graphs with tuneable clustering \cite{miller2009percolation,NewmanRandomClustering}. Our main contributions are summarized below. 
 
	Our first key result characterizes the \emph{probability of the emergence}, i.e., the probability that a spreading process initiated by an infective seed node, selected uniformly at random, leads to an unbounded chain of infections, thus infecting a strictly positive fraction of individuals in the limit of large network size. Next, we derive the critical \emph{epidemic threshold}, thereby defining the boundary of the region in the parameter space inside which only a finite chain of transmissions are observed with high probability and outside which epidemics occur with a positive probability. The analysis becomes particularly challenging as compared to single-strain models\cite{miller2009percolation,NewmanRandomClustering} since characterizing the branching process analysis requires a careful consideration of the infections passed onto node pairs at the endpoints of a triangle. Additionally, we provide a heuristic to estimate the conditional mean of the fraction of infected individuals given that an epidemic outbreak has occurred. We provide extensive simulations validating our theoretical results for the probability of emergence and epidemic threshold, as well as the heuristic for the epidemic size for different mutation and clustering patterns. For {doubly Poisson} contact networks, we observe that clustering increases the threshold of epidemics but reduces the probability of emergence and mean epidemic size. Moreover, through an analytical case study for \emph{one-step irreversible} mutation patterns, we observe that clustering can provide additional pathways for mutations, thereby altering the course of an epidemic. Our results shed light on the epidemic characteristics of multi-strain spreading on clustered networks, paving the way for assessing the risks associated with the emergence of epidemic and information outbreaks. 
		\subsection{Organization} We describe the clustered random network model and the multi-strain transmission model in Section \ref{sec:setup}. In Section~\ref{sec:res-discuss}, we provide our main analytical and experimental results characterizing the emergence of epidemics. We summarize our findings and conculde with future directions in Section~\ref{sec:conc}. A brief outline of the proofs is provided in Section~\ref{sec:proof-sketch}, with further details provided in Section~\ref{sec:proof-details}.
	\section{Problem Setup}
	\label{sec:setup}
	\subsection{Network Model}
	We consider a generalization\cite{miller2009percolation,NewmanRandomClustering} to the standard \emph{configuration model} \cite{ne:newman2018networks,MolloyReed} that generates random graphs with arbitrary degree distribution and tuneable clustering. Note that we could quantify the level of clustering associated with a network in different ways, but here we focus on the notion of {\em global clustering coefficient} defined as
	\begin{equation}\nonumber
		C_{\mathrm{global}} = \frac{3 \times \text{ number of triangles in the network}}{\text{number of connected triples}}
	\end{equation}
	where a connected triple means a single vertex connected by edges to two others.
	The algorithm used to generate random graphs with clustering works as follows. We first specify the probability that an arbitrary node has $s$ single-edges and is part of $t$ triangles through the joint degree distribution $\left\{ q_{s,t} \right\}_{s,t=0}^\infty$. Note that if a node has $s$ single-edges and is part of $t$ triangles, then its degree is $s+2t$ since each triangle adds two edges connecting the node to the other end nodes of the triangle. 
	We can view $s$ as the number of single stubs and $t$ as the number of corners of triangles. In order to create the network, we choose pairs of single stubs uniformly at random and join them to make a complete edge between two nodes, and we choose trios of corners of triangles at random and join them to form a triangle. The total degree distribution in the network is obtained through the joint distribution of single-edges and triangles $\left\{ q_{s,t} \right\}_{s,t=0}^\infty$ as follows,
	\begin{equation} \nonumber
		p_k = \sum_{s,t} q_{s,t} \delta_{k, s+2t}
	\end{equation}
	where $p_k$ denotes the probability that an arbitrary node is of degree $k$ and $\delta_{ij}$ is the Kronecker delta function. For more details on algorithmic construction for the clustered configuration model, we refer the reader to \cite{clustering-vaccination-math,miller2009percolation,NewmanRandomClustering}. In contrast to the standard configuration model, where $C_{\mathrm{global}}$ approaches zero in the limit of large network size {\cite{ne:newman2018networks}}, the quantity $C_{\mathrm{global}}$ is positive for networks generated according to the above algorithm implying the existence of a non-trivial clustering in the network. {We note that the network can admit cycles comprising single-edge and triangle-edges, but as long as the second moments of the distribution of triangle and single-edges is bounded, they occur with a vanishingly small probability in the limit of large network size\cite{clustering-vaccination-math}.
 \begin{figure}[t]
    \centering
    \includegraphics[width=0.33\textwidth]{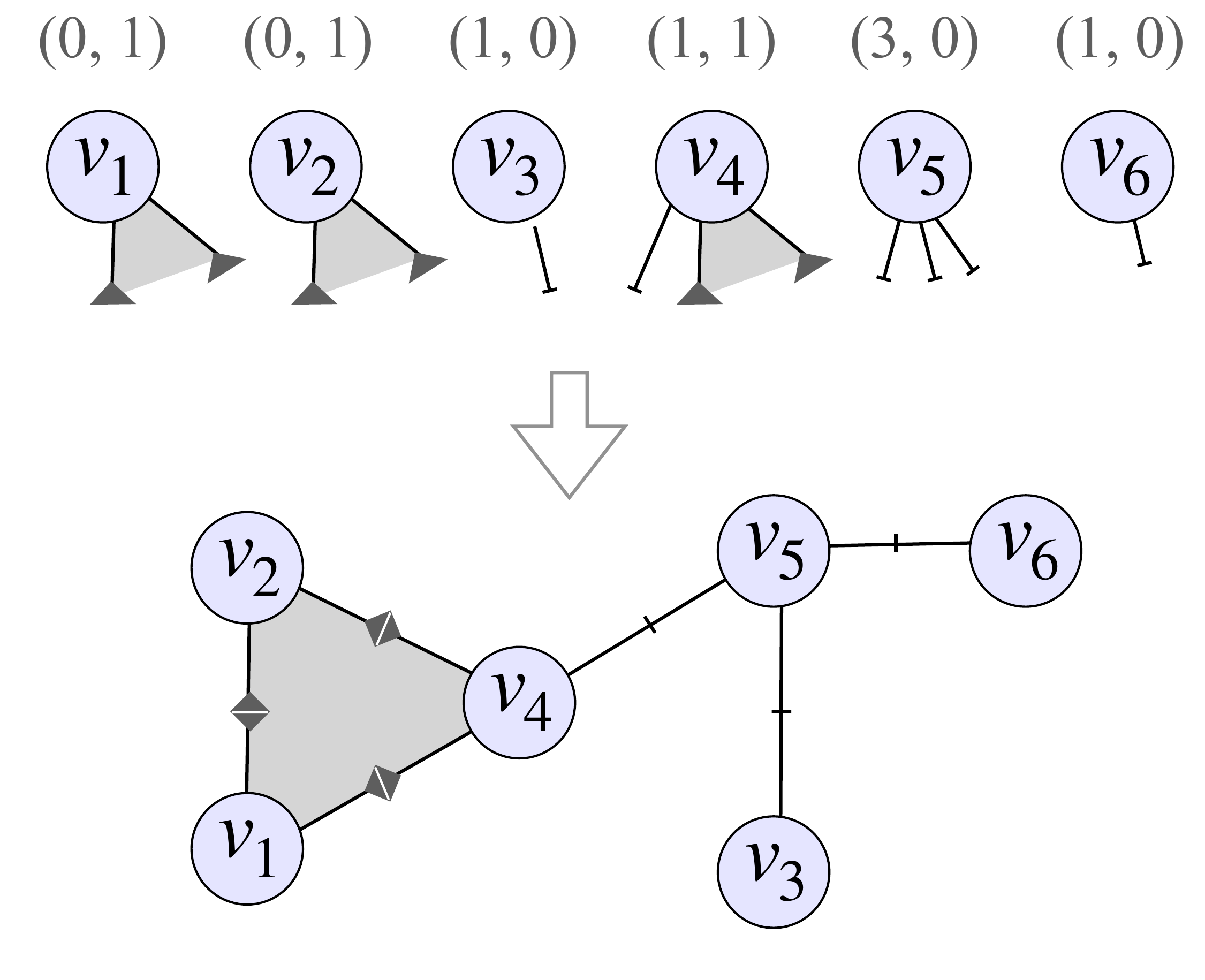}
    \caption{\sl An illustration of the modified configuration model \cite{NewmanRandomClustering} with a given realization for the degree sequence of the single-edges and triangle-edges. We assign single stubs and triangle corners as per the degree sequence, e.g., node $v_1$ admits no single stubs and only one triangle corner. In order to create the network, we choose pairs of single stubs uniformly at random and join them to make a complete edge between two nodes, and we choose triplets of triangle corners uniformly at random and join them to form a triangle. The graph generation algorithm enables generating networks with tuneable clustering coefficients.}
    \label{fig:triangle-config-model-ex}
\end{figure}
		
		\subsection{Transmission Model}
		For modeling the spread of the contagion, we invoke the \emph{multi-strain} model \cite{ne:alexday,ne:pnas,multilayermultistrain} that approximates the spreading dynamics with a {multi-type} branching process \cite{mode1971multitype,mode1971multitype}. Specifically, we allow each \emph{strain} of the contagion to be associated with varying risks of transmission. For $i=1,2,\cdots,m$, we let  $T_i$ denote the \emph{transmissibility} of strain-$i$, i.e., the probability that an infectious node carrying strain-$i$ infects its neighbor. We account for {evolutionary adaptations} or \emph{mutations} in the contagion by specifying the probability $\mu_{ij}$ that strain-$i$ mutates to strain-$j$ within a host, where $i,j=1,2,\cdots,m$ and $\sum_j \mu_{ij} = 1$. The epidemiological and evolutionary processes are assumed to occur on a similar timescale, and each new infection offers an opportunity for mutation \cite{ne:alexday}. We focus on the case where $m=2$, i.e., there are two strains propagating in the population, yet extending our theory to the general case of $m$ strains is straightforward. The transmissibility and mutation probabilities for different strains are encoded through the transmission and mutation matrices, respectively denoted as $\pmb{T}$ and $\pmb{\mu}$ below.	\begin{align}
			\pmb{T}= \left[    \begin{matrix}
				T_1 &  0\\ 
				0 &  T_2
			\end{matrix} \right], \nonumber  \qquad 
			\pmb{\mu}= \left[   \begin{matrix}
				\muoneone &  \muonetwo\\ 
				\mutwoone &  \mutwotwo
			\end{matrix} \right]. \nonumber 
		\end{align}

		We consider a multi-type branching process \cite{ne:alexday} that starts by selecting a node uniformly at random and infecting it with a particular strain, then exploring all the susceptible neighbors that can be reached and infected due to this node (Figure~\ref{fig:0}). We assume that initially all nodes are in the susceptible state. Further, as per \cite{ne:alexday}, we assume that once an individual recovers from an infection with either strain, they cannot be reinfected with any strain. At each stage in the process, a node carrying strain-$i$ infects its neighbors independently with probability $T_i$. It stays active for one round, after which it turns inactive. This corresponds to an SIR (Susceptible-infectious-recovered) epidemic where some nodes remain susceptible throughout because the epidemic never reaches them, while others get infected and effectively recover after one round of transmissions. We assume that co-infection with multiple strains is not possible and subsequent to each transmission event, the contagion mutates to strain $j$ with probability $\mu_{i,j}$, where $i,j=1,2,\cdots,m$ within the host. The process continues recursively until no further infections are possible. Further details regarding the multi-strain spreading process are presented in Section~\ref{subsec:sim}. 

\subsection{Metrics of Interest}
We characterize an outbreak as an \emph{epidemic} if the introduction of the contagion to a host population causes an outbreak infecting a positive fraction of individuals in the limit of large network size. In contrast, we characterize outbreaks as being \emph{self-limited} when the spreading process dies out after a finite number of transmission events. We study the following metrics \cite{ne:newman2018networks}, which are used in quantifying and assessing risks during the early stages of an outbreak.\\ i) The \emph{probability of emergence} starting from strain-$i$ is the probability that the spreading process initiated by a seed node chosen uniformly at random, carrying strain-$i$ infects a positive fraction of the population in the limit of large network size, i.e., triggers an outbreak of size $\Omega(n)$. \\ ii) The \emph{epidemic threshold} defines a boundary of the region inside which the outbreak always dies out after infecting only a finite number of individuals, while outside which an epidemic outbreak occurs with a positive probability. \\
iii) The \emph{mean epidemic size} is defined as the conditional mean of the fraction of infected individuals given that an epidemic outbreak has occurred.

		\section{Results and Discussion}
		\label{sec:res-discuss}
		\subsection{Preliminaries} \label{subsec:res-prelim}The analysis of the probability of emergence relies on recursive equations linking the number of nodes infected by the \emph{seed} node to the number of nodes consequently infected by \emph{later-generation} infectives. To enable such a recursive analysis, we first present preliminary facts regarding the possible configurations based on the type of strain acquired by endpoints of a triangle emanating from an infectious parent node strain-$i$. A graphical illustration of the resulting configurations for the case when the triangle emanates from a parent node carrying the type-$1$ strain (indicated in blue) is given in Figure~\ref{clustEvo_fig:fig1}. We present the corresponding probability of each configuration $(p_{ij})$ in Table~\ref{table:tab2}, with $j = {1,2,\cdots,6}$ and where $i=1,2$ corresponds to the type of strain carried by the parent node. 
		\begin{figure}[htbp]
			\centering
			\includegraphics[scale=0.135]{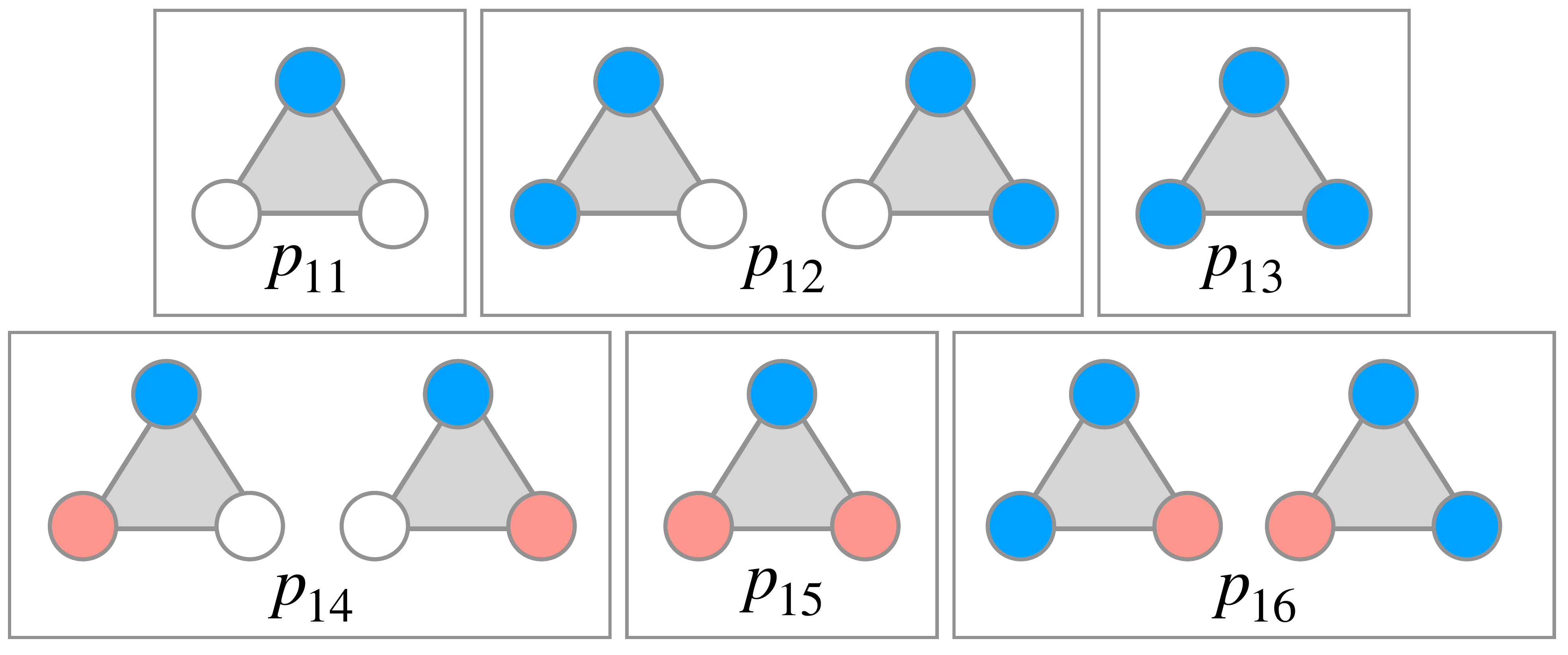}
			\caption{\sl Different possible configurations for a triangle emanating from a parent node carrying strain-$1$. Nodes that acquire strain-$1$ (resp., strain-$2$) after mutation are indicated in blue (resp., red).  The configurations are based on whether the node at either endpoint of the triangle gets infected and the resulting strain it acquires after mutation.} 
			\label{clustEvo_fig:fig1}
		\end{figure} 
		\begin{table}[htbp]
			\vspace{2mm}
			\centering
			\begin{tabular}{|l|c|}
				\hline
				$p_{i1}$ &$\left( 1- T_i \right)^2$ \\ \hline
				$p_{i2}$ &$2 T_i \mu_{i1}\left(1-T_i \right) \left(1-T_1 \right)$ \\ \hline
				$p_{i3}$ &$\left(T_i \mu_{i1} \right)^2 + 2T_i\mu_{i1}\left( 1- T_i \right) T_1 \mu_{11}$ \\ \hline
				$p_{i4}$ &$2 T_i \mu_{i2} \left(1-T_i \right) \left(1-T_2 \right)$ \\ \hline
				$p_{i5}$ &$\left(T_i \mu_{i2} \right)^2 + 2T_i\mu_{i2}\left( 1- T_i \right) T_2 \mu_{22}$ \\ \hline
				$p_{i6}$ &$2 \left( T_i^2 \mu_{i1} \mu_{i2} + T_i \mu_{i1} \left(1-T_i \right)T_1 \mu_{12} +  T_i \mu_{i2} \left(1-T_i \right)T_2 \mu_{21}  \right)$ \\ \hline 
			\end{tabular}
			\caption{\sl The probability $p_{ij}$ of occurrence for each of the scenarios in Figure~\ref{clustEvo_fig:fig1}, where $i=1,2$ corresponds to the strain carried by the parent and $j=1,\dots,6$ corresponds to the configuration at the endpoints of the triangle emanating at the parent node.  }
			\label{table:tab2}
		\end{table}
	We illustrate how the probabilities are derived for a sample configuration $p_{13}$ in Figure~\ref{clustEvo_fig:fig1} and direct the reader to Section~\ref{sec:proof-details} for derivations of the remaining scenarios in Table~\ref{table:tab2}.
In Figure~\ref{clustEvo_fig:fig1}, the configuration corresponding to $p_{13}$ occurs when i) the parent node infects both endpoints of the triangle, and they acquire strain-$1$, {\em or} ii) the parent node infects one of the two endpoints but fails to infect the other endpoint which later gets infected. Its easy to verify that $p_{13}=\left(T_1 \mu_{11} \right)^2 + 2T_1\mu_{11}\left( 1- T_1 \right) T_1 \mu_{11}$, where the factor $2$ is due to symmetry.} Now that we have established the framework for presenting our main results, we present our first analytical result characterizing the probability of emergence.

		\subsection{Probability of Emergence}

		\begin{theorem}[Probability of Emergence]
			{\sl For a multi-strain branching process with parameters $(\pmb{T}, \pmb{\mu})$, initiated by a randomly selected seed node carrying strain-$i$, on a clustered network with a given joint degree distribution of single-edges and triangles ($q_{s,t}$), for $i=1,2$, we have
				\begin{align}
					\pr {[\rm Emergence]} =1- \sum_{s,t} q_{s,t} (h_i(1))^s (g_i(1))^t,			
				\end{align}
				where $h_i(1), g_i(1)$ are the smallest non-negative roots of the fixed point equations:
				\begin{align}
					h_i(1) &= 1-T_i+T_i  \Big( \mu_{i1} \sum_{s,t} \frac{s q_{s,t}}{\langle s \rangle} h_1(1)^{s-1} g_1(1)^t  \nonumber\\&\quad+  \mu_{i2} \sum_{s,t} \frac{s q_{s,t}}{\langle s \rangle} h_2(1)^{s-1} g_2(1)^t \Big),	\label{eq:ne:rec1}\\
					g_i(1)  
					&=p_{i1} + p_{i2}  \left(  \sum_{s,t} \frac{t q_{s,t}}{\langle t \rangle} h_1(1)^s g_1(1)^{t-1}\right)\nonumber\\
					&\quad+ p_{i3} \left(  \sum_{s,t} \frac{t q_{s,t}}{\langle t \rangle} h_1(1)^s g_1(1)^{t-1} \right)^2\nonumber\\& \quad+  p_{i4}  \left(  \sum_{s,t} \frac{t q_{s,t}}{\langle t \rangle} h_2(1)^s g_2(1)^{t-1} \right)\nonumber\\
					& \quad+ p_{i5} \left(  \sum_{s,t} \frac{t q_{s,t}}{\langle t \rangle} h_2(1)^s g_2(1)^{t-1} \right)^2 \nonumber\\&  \quad +p_{i6} \left(  \sum_{s,t} \frac{t q_{s,t}}{\langle t \rangle} h_1(1)^s g_1(1)^{t-1} \right)\nonumber \\
					&  \qquad\cdot \left(  \sum_{s,t} \frac{t q_{s,t}}{\langle t \rangle} h_2(1)^s g_2(1)^{t-1} \right) \label{eq:ne:rec2},~ i=1,2.
			\end{align}}
			\label{thm:prob}
		\end{theorem}
	We provide an outline for the proof of Theorem~\ref{thm:prob} in Section~\ref{subsec:thm-proof} and additional details in Section~\ref{sec:proof-details}.   We present numerical simulations in Section~\ref{subsec:sim}. 
		
		\subsection{Epidemic Threshold}
		Our following result characterizes the epidemic threshold.
		\begin{theorem}[Epidemic Threshold]
			{\sl For a multi-strain branching process with parameters $(\pmb{T}, \pmb{\mu})$ on a clustered network, we define
				\begin{align}
					\pmb{J}&= 	\left[
					\begin{matrix}
						\pmb{\Pi}&\pmb{0} \\
						\pmb{0} & \pmb{\Delta}
					\end{matrix}
					\right]
					\left[
					\begin{matrix}
						\frac{\langle s^2 \rangle - \langle s \rangle}{\langle s \rangle} 	\pmb{\rm I}&\frac{\langle st \rangle}{\langle s \rangle}  \pmb{\rm I}\\
						\frac{\langle st \rangle}{\langle t \rangle}   \pmb{\rm I}& \frac{\langle t^2 \rangle - \langle t \rangle}{\langle t \rangle} \pmb{\rm I}
					\end{matrix}
					\right]. \label{eq:JacobianMat}
				\end{align}
				Let
				$\rho(\boldsymbol{J})$ denote the spectral radius of $\rho(\boldsymbol{J})$. The epidemic threshold is given by $\rho(\boldsymbol{J})=1$, where
				\begin{align}
					\pmb{\Pi}&= \left[   \begin{matrix}
						T_1\muoneone &  T_1\muonetwo\\ 
						T_2\mutwoone &  T_2\mutwotwo
					\end{matrix} \right], \nonumber \\
					\pmb{\Delta}&= \left[    \begin{matrix}
						p_{12} + 2p_{13}+p_{16} &  p_{14}+ 2p_{15} + p_{16} \\ 
						p_{22}+ 2p_{23}+p_{26} &  p_{24}+ 2p_{25}+ p_{26}
					\end{matrix} \right]. \label{eq:pianddelta}
				\end{align}
			}\label{thm:threshold}
		\end{theorem}
		Observe that for $i,j=1,2$, the matrix entry $\pmb{\Pi}_{ij}$ corresponds to the probability that a parent node carrying strain-$i$ infects its neighbor via a single-edge with strain-$j$ after mutation. Similarly, $\pmb{\Delta}_{ij}$ corresponds to the mean number of nodes that acquire strain-$j$ at the endpoints of a triangle emanating from a parent node carrying strain-$i$. We note that the epidemic threshold as presented in Theorem~\ref{thm:threshold} is a strict generalization of the multi-strain model without clustering, the threshold for which can be inferred by substituting $\pmb{\Delta}=0$ in \eqref{eq:JacobianMat}. In Section \ref{subsec:contrast} we further decompose the epidemic threshold to delineate the impact of clustering through a case-study with irreversible mutations. A brief outline for proving Theorem~\ref{thm:threshold} is provided in Section~\ref{subsec:proof-thresh}; the full derivation is provided in Section~\ref{subsec:additional-thm-thresh}. 
The above result states that the epidemic threshold is tied to the spectral radius of the Jacobian matrix $J$, i.e., if $\rho(\boldsymbol{J}) > 1$ then an epidemic occurs with a positive probability, whereas if $\rho(\boldsymbol{J}) \leq 1$ then with high probability the infection causes a \emph{self-limited} outbreak with fraction of infected nodes going to 0 as $n \rightarrow \infty$.
\subsection{Epidemic Size}
For the case when $\rho(\boldsymbol{J}) > 1$, which is known as the \emph{super critical} regime, we know that an epidemic occurs with a positive probability. A natural metric of interest in this scenario is the conditional mean of the fraction of infected individuals given an epidemic outbreak, referred to as the mean epidemic size. In the literature on multi-strain contagions \cite{ne:alexday
,ne:pnas
,multilayermultistrain}, it is known that above the epidemic threshold, the mean epidemic size differs from the probability of emergence. Furthermore, calculating the mean epidemic size through expensive mean-field approximations involves fixed-point iterations with each function evaluation entailing nested infinite summations, resulting in high computational costs \cite{ne:pnas,multilayermultistrain}. Inspired by the effectiveness of bond percolation in predicting size \cite{kenah2007second,miller2007epidemic,ne:pnas
,multilayermultistrain}, as an efficient heuristic for computing the mean epidemic size, we propose transforming the multi-strain contagion into a single contagion with transmissibility $\Tilde{T}$ defined by the spectral radius of the matrix $\Pi$, i.e., 
\begin{align}
    \Tilde{T}=\rho(\pmb{\Pi}). \label{eq:single-strain-tx}
\end{align}
 For multi-strain spreading on clustered networks, we predict the mean epidemic size using our proposed transformation $\rho(\pmb{\Pi})\rightarrow \Tilde{T}$ and we leverage existing results on the mean epidemic size for single-strain spreading on clustered networks \cite{NewmanRandomClustering}.  In the succeeding section, we provide an empirical performance evaluation of our heuristic solution for epidemic sizes under various transmission and mutation patterns, demonstrating its effectiveness in predicting the epidemic size.

\begin{figure}[t]
			\centering
			\includegraphics[scale=0.23]{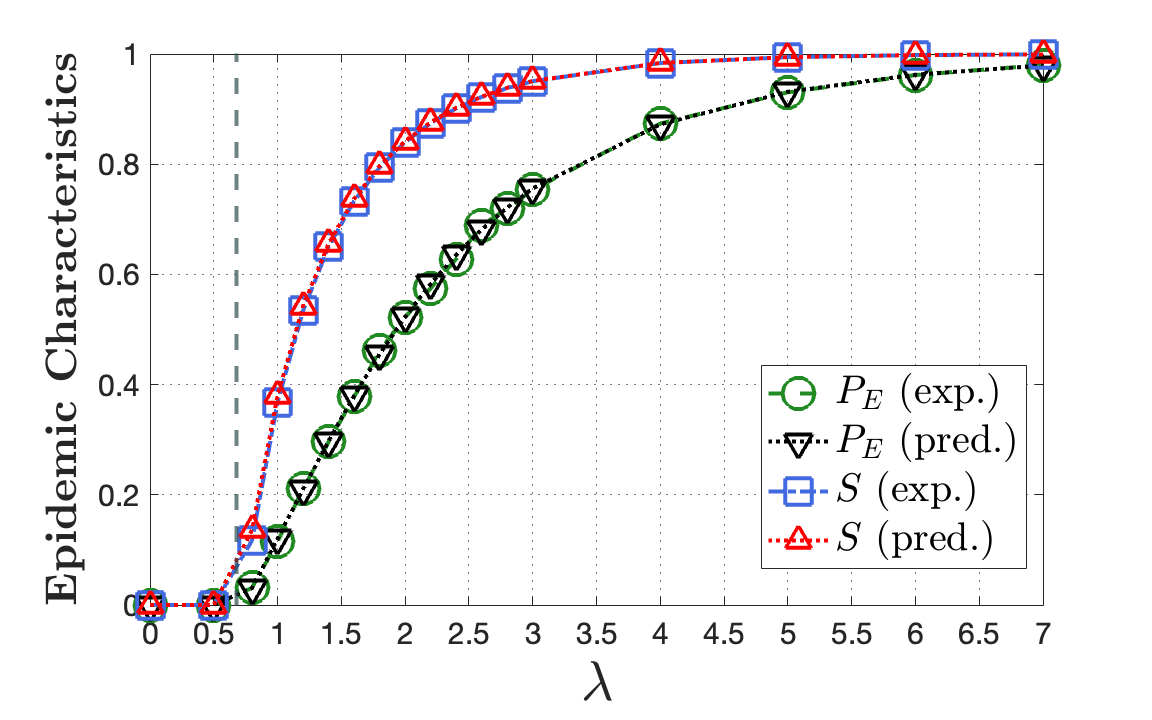}
			\caption{\sl The probability of emergence on contact networks with doubly Poisson distribution \eqref{eq:doublypoisson}, with the distribution for single-edges and triangles, respectively parameterized by $\lambda_s$ and $\lambda_t$. For $\lambda_s = \lambda_t = \lambda$, we vary $\lambda$ in the interval $(0,10)$. The analytically predicted probability (indicated as $P_E$(pred.)) and epidemic threshold (indicated by the vertical dashed line corresponding to $\rho( \pmb{J})=1$) are derived from Theorems~\ref{thm:prob} and \ref{thm:threshold}, respectively, while the mean epidemic size ($S$(pred.)) computed by the proposed transformation to a single-strain model using \eqref{eq:single-strain-tx}. The empirical probability of emergence ($P_E$(exp.)) and the conditional mean epidemic size ($S$(exp.)), given that it occurs averaged over $1.5\times10^4$  independent experiments. For each data point, the network size $n$ is $2 \times 10^5$, the number of independent experiments is $1.5\times10^4$, and we set $T_1=0.2, T_2=0.5$ and $\mu_{11}=\mu_{22}=0.75$.   }
			\vspace{-4mm}
			\label{fig:sim1}
		\end{figure}

Next, we describe intuitively why our proposed heuristic is a reasonable transformation to a single-strain model. We consider transforming the multi-strain spreading process with two strain types to a multi-type \cite{mode1971multitype} branching process with progeny types-$\{a,b\}$, with $a,b =1,2$, where the index $a$ corresponds to the strain that a susceptible individual gets infected by and the index $b$ corresponds to the strain acquired after mutation. For example, if a susceptible individual is infected by an infectious neighbor carrying strain-1, and a subsequent mutation leads to strain-2 within this host individual, we refer to them as a progeny of type $\{1,2\}$. For notational simplicity, we can define types-$\{1,1\},\{1,2\},\{2,1\}$, and $\{2,2\}$ respectively as type-$1,2,3$, and $4$. Let $\beta$ denote the excess mean degree distribution. Next, we can define a mean matrix $\pmb{M} \in \mathbb{R}^{4\times4}$, where $\pmb{M}_{ij}$ denotes the mean number of type-$j$ progenies produced by a type-$i$ parent:
 \begin{align}\pmb{M}=
\beta\begin{bmatrix}
T_1 \mu_{11} & T_1 \mu_{12} & 0 & 0 \\
0 & 0 & T_2 \mu_{21} & T_2 \mu_{22} \\
T_1 \mu_{11} & T_1 \mu_{12} & 0 & 0 \\
0 & 0 & T_2 \mu_{21} & T_2 \mu_{22}
\end{bmatrix}. \label{eq:sizepredmatrix}
\end{align}
It can be verified that the spectral radius of $\pmb{M}/\beta$ is identical to $\pmb{\Pi}$, thus motivating our choice of heuristic. However, we note that the heuristic above, which effectively treats the multi-strain process as bond percolation with effective percolation probability $\rho(\pmb{\Pi})$, is unable to capture the probability of emergence. This can be attributed to the observation that for multi-strain spreading, the infection events are conditionally independent given the type of strain carried by a node and dependent otherwise. Our proposed transformation \eqref{eq:single-strain-tx} tackles the challenge of conditional dependence of infection events by subsuming the mutation event in defining the progeny type. Further, the transformation matches the \emph{mean} number of progenies of each type, and we observe that it is able to predict the epidemic size (Figures~\ref{fig:sim1}-\ref{fig:eqtxcase}) but fails to capture the probability of emergence, which we analyze in Theorem~\ref{thm:prob}.



		\subsection{Simulations}
		\label{subsec:sim}
		Next, we describe the simulation setup for simulating the multi-strain spreading on a clustered network. Unless stated otherwise, the spreading process is initiated by selecting a \emph{seed} node uniformly at random and infecting it with strain-$1$. The seed node infects each neighbor independently with probability $T_1$, following which, each newly infected neighbor mutates independently to strain-$2$ with probability $\mu_{12}$. In the $k^{\rm th}$ round, a node that was infected by a node carrying strain-$i$ first undergoes mutation with probability given through the mutation matrix before attempting to infect its neighbors during the $k+1^{\rm th}$ round. Each node is assumed to be infectious only for one round. As the infections continue to grow, both strains might co-exist in the population. Moreover, the presence of cycles in the contact network can simultaneously expose a susceptible node to multiple infections. We assume that co-infection is not possible, and resolve the exposure to multiple infections as follows. If a node is exposed to $x$ infections of strain-$1$ and $y$ infections of strain-$2$ simultaneously, the node becomes infected with strain-$1$ (respectively, strain-$2$) with probability $x/(x+y)$ (respectively, $y/(x+y)$) for any non-negative constants $x$ and $y$. The process terminates when no further infections are possible.
  
		The underlying contact network is modeled by random graphs with clustering, where the joint degree sequence $q_{s,t}$ is given by the {\em doubly Poisson distribution}, i.e., the number of single-edges and triangles are independent, and they follow a Poisson distribution.
		Namely, we set
		\begin{equation}
			q_{s,t} = e^{-\lambda_s} \frac{\left( \lambda_s \right)^s }{s!} \cdot e^{-\lambda_t} \frac{\left( \lambda_t \right)^t }{t!}, \quad s,t=1,\ldots \label{eq:doublypoisson}
		\end{equation}
		with $\lambda_s$ and $\lambda_t$ denoting the mean number of single-edges and triangles, respectively. Throughout we set the network size $n$ to $2 \times 10^5$ and we use a fractional size of $0.05$ as the threshold for the emergence of an epidemic outbreak. Further, we perform $1.5 \times 10^4$ independent experiments for each data point. The code used for the experiments is available in \cite{code-details-2024}.

\begin{figure*}[htbp]
    \centering
    \captionsetup[subfloat]{captionskip=-2pt, position=top} 
    \subfloat[$T_1=0.2, T_2=0.5, \mu_{11}=0.25, \mu_{22}=0.75$\label{1a}]{%
        \includegraphics[width=0.49\linewidth]{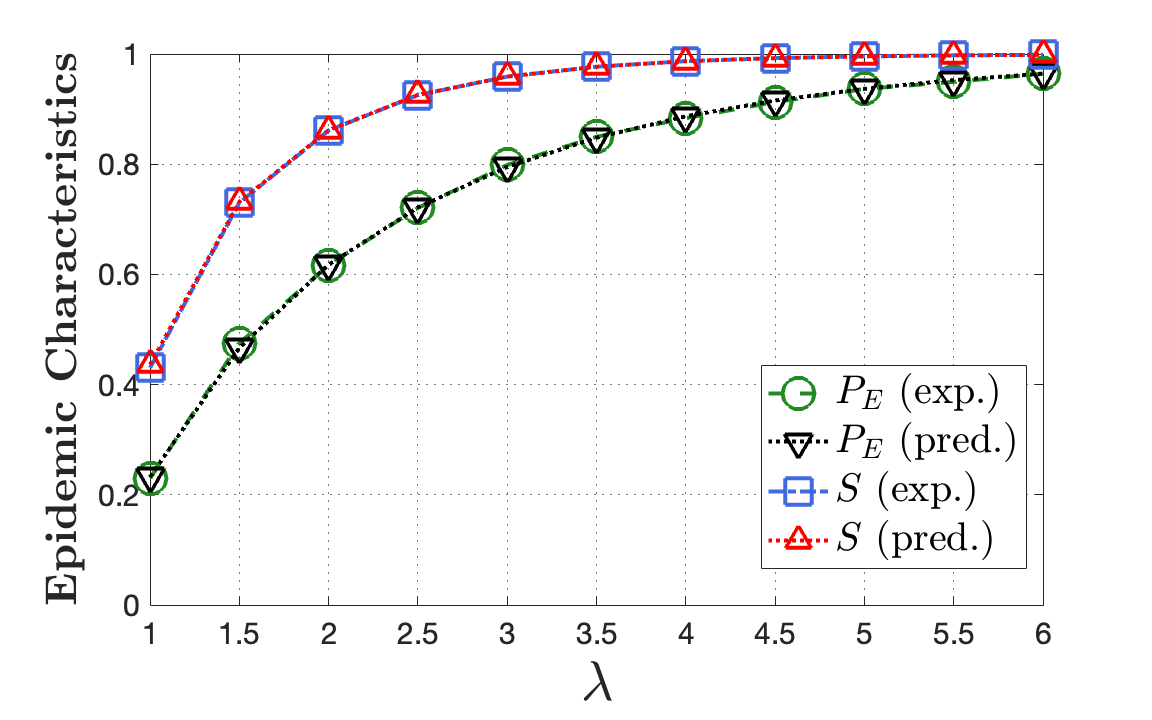}}
    \subfloat[$T_1=0.2, T_2=0.5, \mu_{11}=0.5, \mu_{22}=0.75$\label{1b}]{%
        \includegraphics[width=0.49\linewidth]{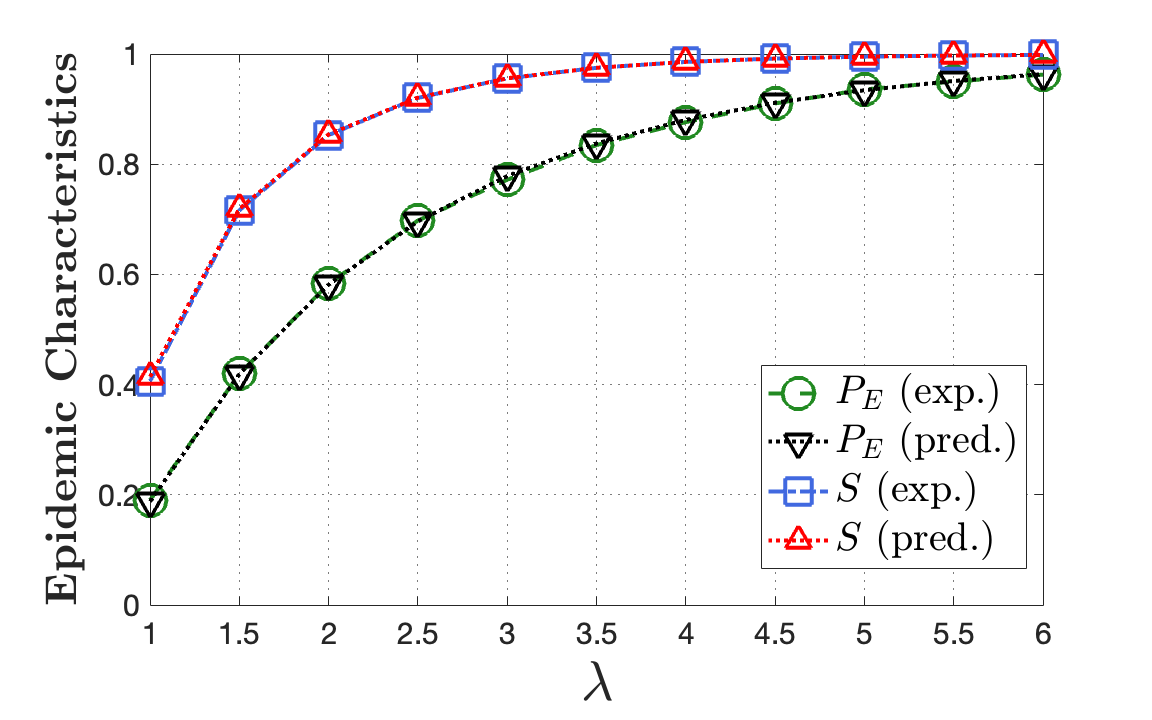}}\\
        \vspace{25pt}
    \subfloat[$T_1=0.2, T_2=0.5, \mu_{11}=0.9, \mu_{22}=0.75$\label{1c}]{%
        \includegraphics[width=0.49\linewidth]{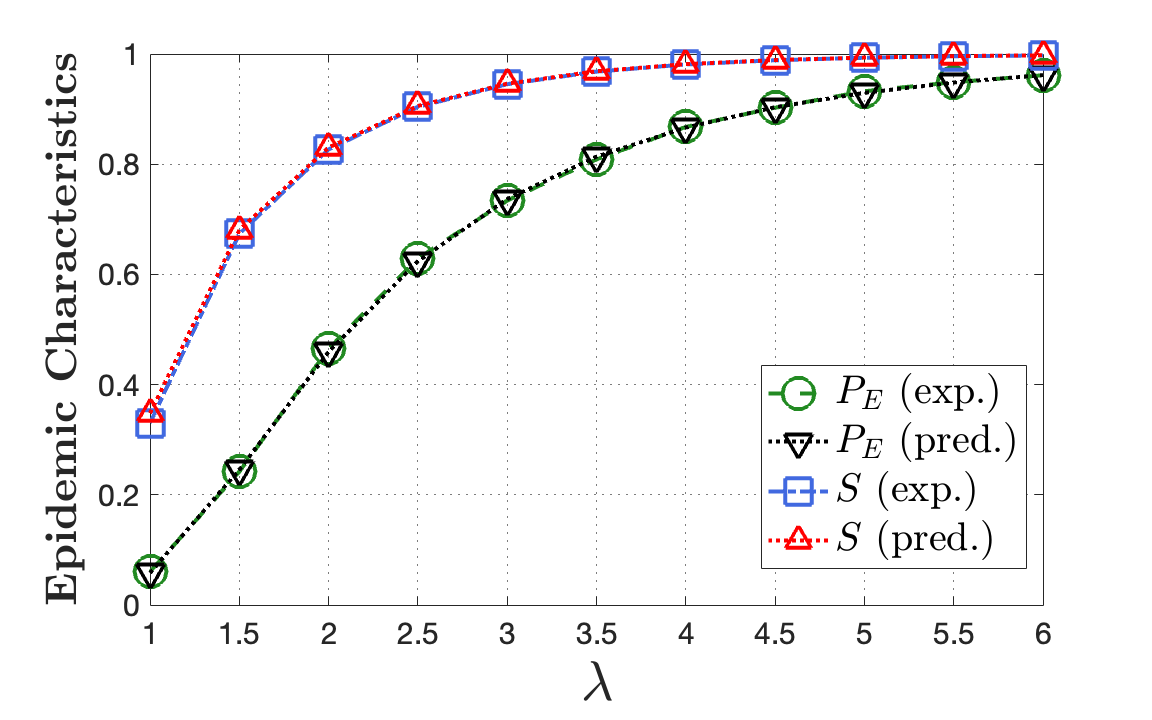}}
    \subfloat[$T_1=0.2, T_2=0.5, \mu_{11}=0.75, \mu_{22}=0.25$\label{1d}]{%
        \includegraphics[width=0.49\linewidth]{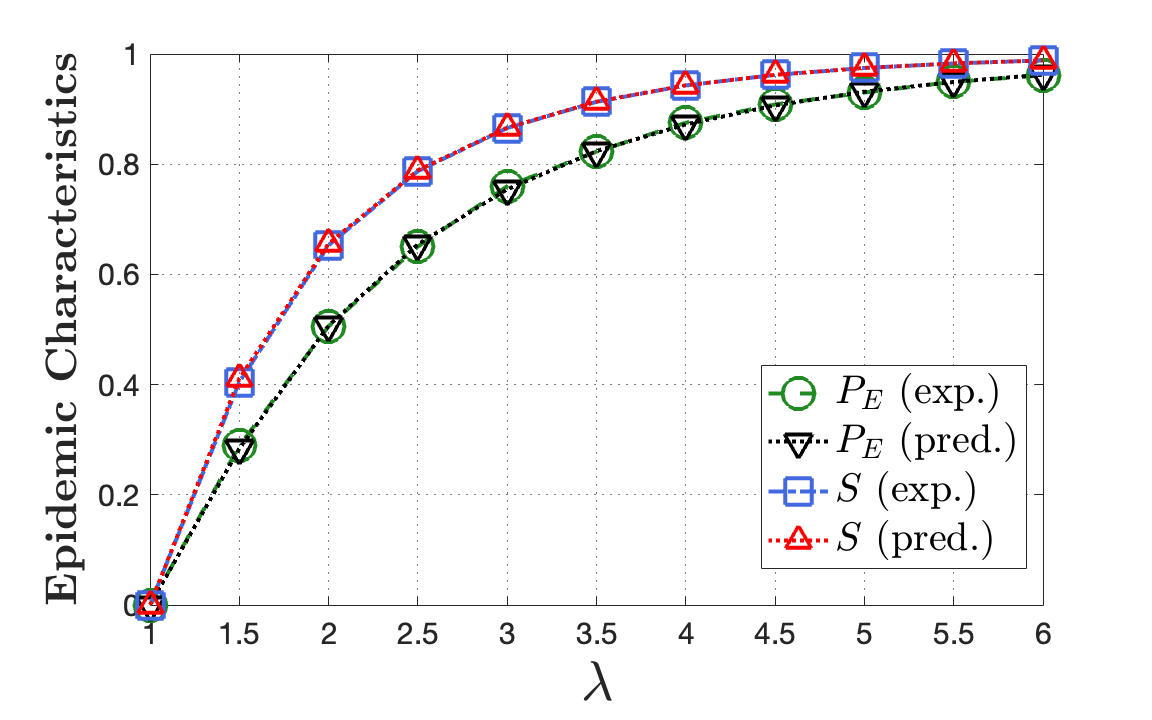}}
        \\         \vspace{25pt}
    \subfloat[$T_1=0.2, T_2=0.5, \mu_{11}=0.75, \mu_{22}=0.5$\label{1e}]{%
        \includegraphics[width=0.49\linewidth]{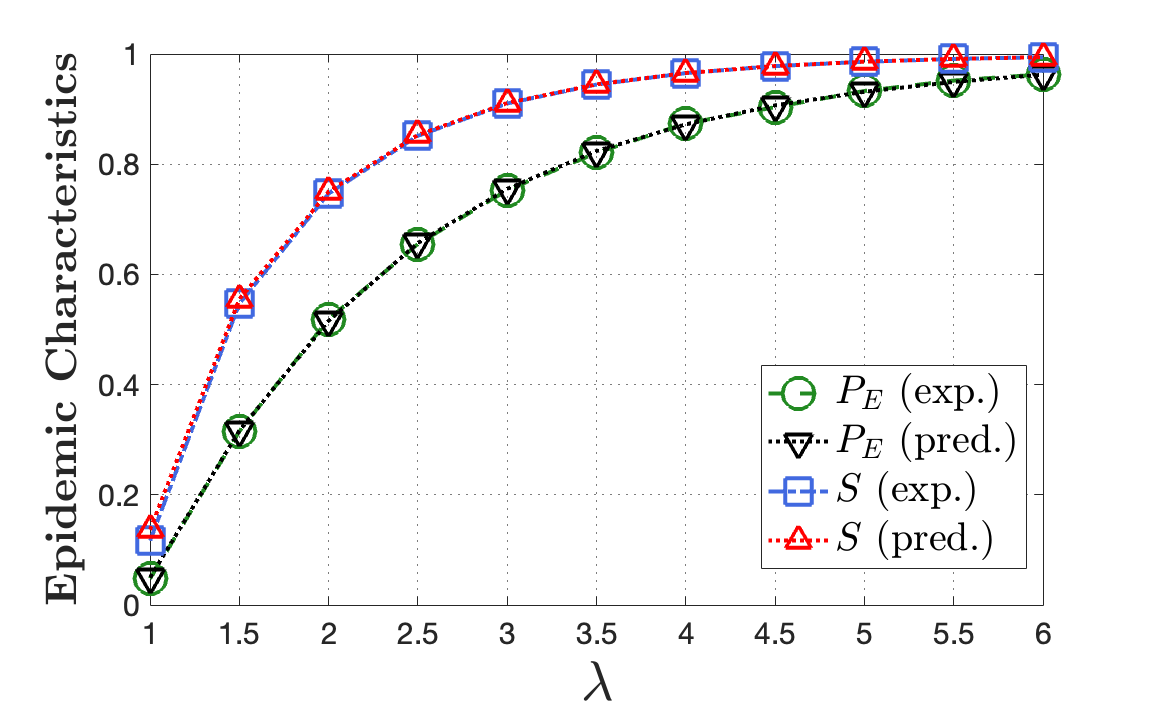}}
    \subfloat[$T_1=0.2, T_2=0.5, \mu_{11}=0.75, \mu_{22}=0.9$\label{1f}]{%
        \includegraphics[width=0.49\linewidth]{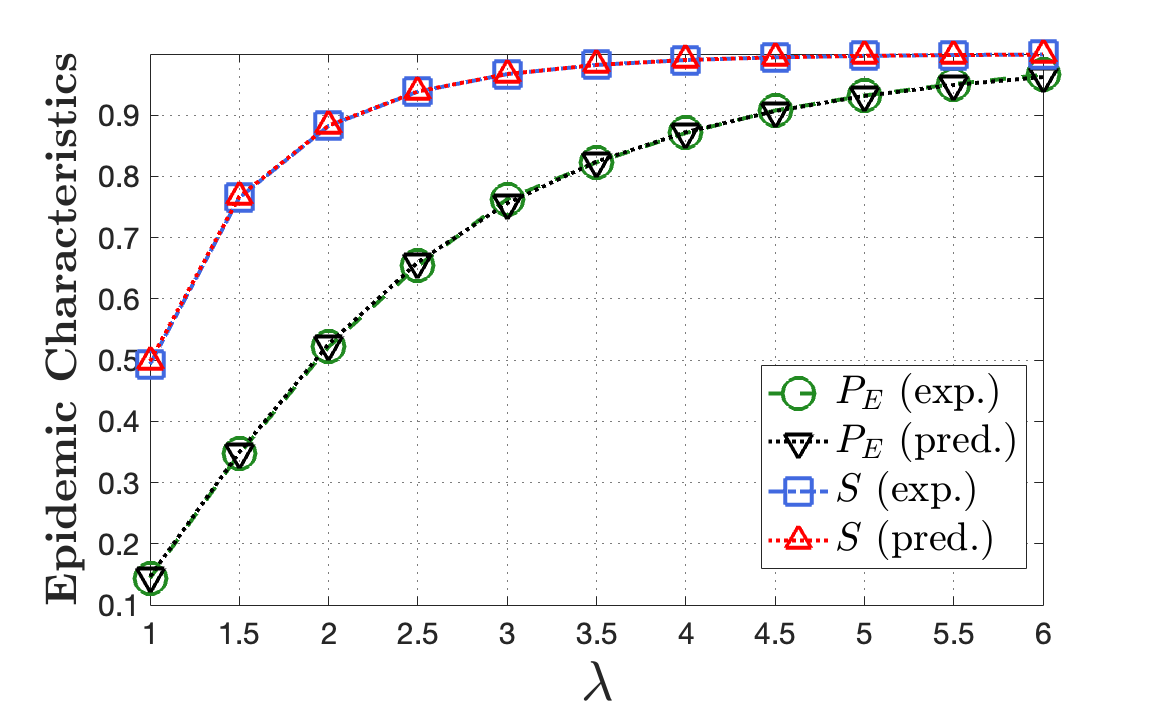}}
    \caption{\sl We plot the probability of emergence and expected epidemic size. We consider doubly Poisson distribution \eqref{eq:doublypoisson} with the distribution for single-edges and triangles, respectively parameterized by $\lambda_s$ and $\lambda_t$, such that $\lambda_s = \lambda_t = \lambda$. The network size $n=2 \times 10^5$, and we report the empirical probability of emergence ($P_E$ (exp.)) and mean epidemic size ($S$ (exp.)) as averaged over $1.5\times10^4$ independent experiments. The analytically predicted probability ($P_E$ (pred.)) is derived from Theorems~\ref{thm:prob} while the mean epidemic size ($S$ (pred.)) is computed by transforming to a single-strain model using \eqref{eq:single-strain-tx}. We let $T_2$ be the more transmissible strain with $T_2=0.5,~T_1=0.2$. Using the parameters in Fig.~\ref{fig:sim1}, as a baseline, through sub-figures (a), (b), and (c), we vary $\mu_{11}\in\{0.25,0.5,0.9\}$ while holding $\mu_{22}$ constant. Conversely, in sub-figures (d), (e), and (f), we hold $\mu_{22}=0.75$ constant and vary $\mu_{22}\in\{0.25,0.5,0.9\}$. We observe good agreement between our predictions for the probability of emergence and mean epidemic size and the corresponding observed values in the simulations.}
    \label{fig:6subfigures}
\end{figure*}

\begin{figure}[htbp]
    \centering
        \includegraphics[width=\linewidth]{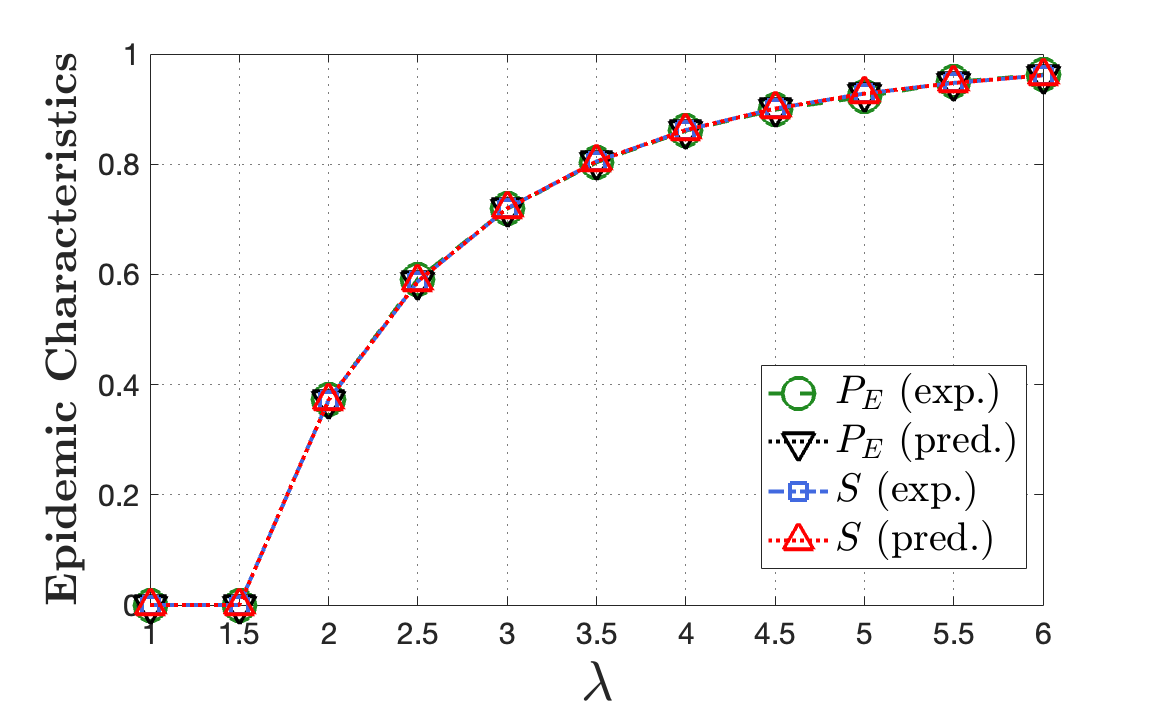}
    \caption{{\sl We set the network parameters parameters same as Figures~\ref{fig:sim1} and ~\ref{fig:6subfigures}. However, we now consider the scenario where strains have identical transmissibility but different mutation probabilities. Specifically, we set $T_1=T_2=0.2$ and $\mu_{11}=0.5<\mu_{22}=0.75$. The  quantities $P_E$(exp.) and $S$(exp.) correspond to the empirical average reported over $1.5\times10^4$ independent experiments. We note that for the case of equal strain transmissibilities, all of the four quantities- the experimentally observed and the predicted probability of emergence and the mean epidemic size coalesce to the same value. }}
     \label{fig:eqtxcase}
\end{figure}

In Figure~\ref{fig:sim1}, we consider the cases when $\lambda_s = \lambda_t = \lambda$ with $\lambda$ varying from $1$ to $10$. We compute the critical value of $\lambda$ for which (\ref{eq:JacobianMat}) has a spectral radius of one, i.e., $\rho( \pmb{J})=1$, to mark the phase transition point. We also plot the empirical probability of emergence ($P_E$ (exp.)) and the mean epidemic size ($S$ (exp.)) averaged over $1.5\times10^4$ independent experiments, and the analytical prediction for the probability of emergence ($P_E$ (pred.)) from Theorem~\ref{thm:prob} and our heuristic estimate for the mean epidemic size ($S$ (pred.)) computed by transforming to a single-strain model using \eqref{eq:single-strain-tx}. We set the transmission parameters $T_1=0.2, T_2=0.5$ and the mutation parameters $\mu_{11}=\mu_{22}=0.75$. We observe that both the probability of emergence and the epidemic size transition from zero to a positive value around the epidemic threshold. Our theoretical results on the probability of emergence and phase transition point are in agreement with simulation results. Next, we further investigate the case of different transmission and mutation matrices in Figures~\ref{fig:6subfigures} and \ref{fig:eqtxcase} and vary $\lambda \in [1,6]$ in increments of 0.5. In Figures \ref{1a}, \ref{1b}, and \ref{1c}, we set $\mu_{11}=0.25,0.5$, and $0.9$ respectively, while keeping the other transmission and mutation parameters constant with $T_1=0.2, T_2=0.5$, and $\mu_{22}=0.75$. Note that since we have set strain-2 to be the more transmissible and the seed node acquires strain-1, we note that the probability of emergence decreases as we increase $\mu_{11}$ since it becomes less likely for a mutation to strain-2 to appear ($\mu_{12}=1-\mu_{11}$) within a host infected with strain-1. Next, in Figures~\ref{1d}, \ref{1e}, and \ref{1f}, we vary $\mu_{22}=0.25,0.5$, and $0.9$, while keeping $T_1=0.2, T_2=0.5$, and $\mu_{11}=0.75$ constant. Throughout, we observe a good agreement between the simulations and the analytical predictions for the probability of emergence from Theorem~\ref{thm:prob} and our heuristic estimate for the mean epidemic size using \eqref{eq:single-strain-tx}. Next, in Figure~\ref{fig:eqtxcase}, we consider the scenario when strains have the same transmissibility but varying mutation probabilities, setting $T_1=T_2=0.2$ and $\mu_{11}=0.5$, $\mu_{22}=0.75$. We note the observed and predicted probabilities of emergence, along with the mean epidemic size, converge to a single value in line with the observation for other multi-strain models \cite{multilayermultistrain}.


\begin{figure*}[t]
    \centering
    \captionsetup[subfloat]{captionskip=-2pt, position=top} 
{%
        \includegraphics[width=0.49\linewidth]{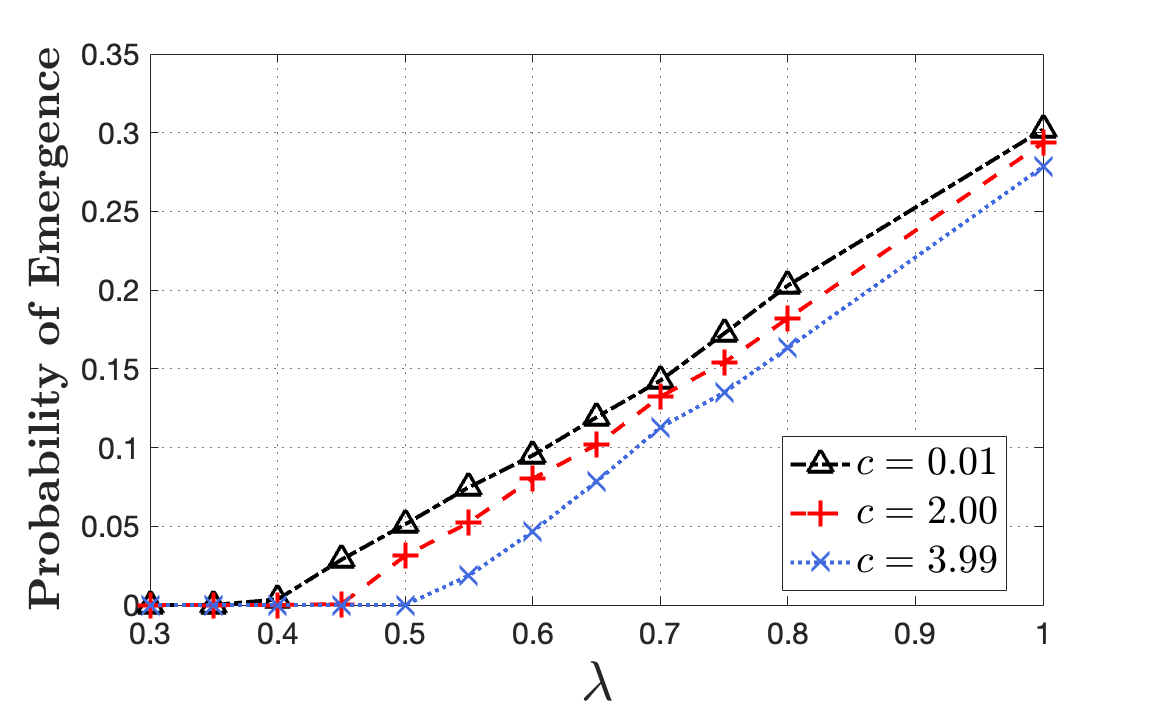}}
 {%
        \includegraphics[width=0.49\linewidth]{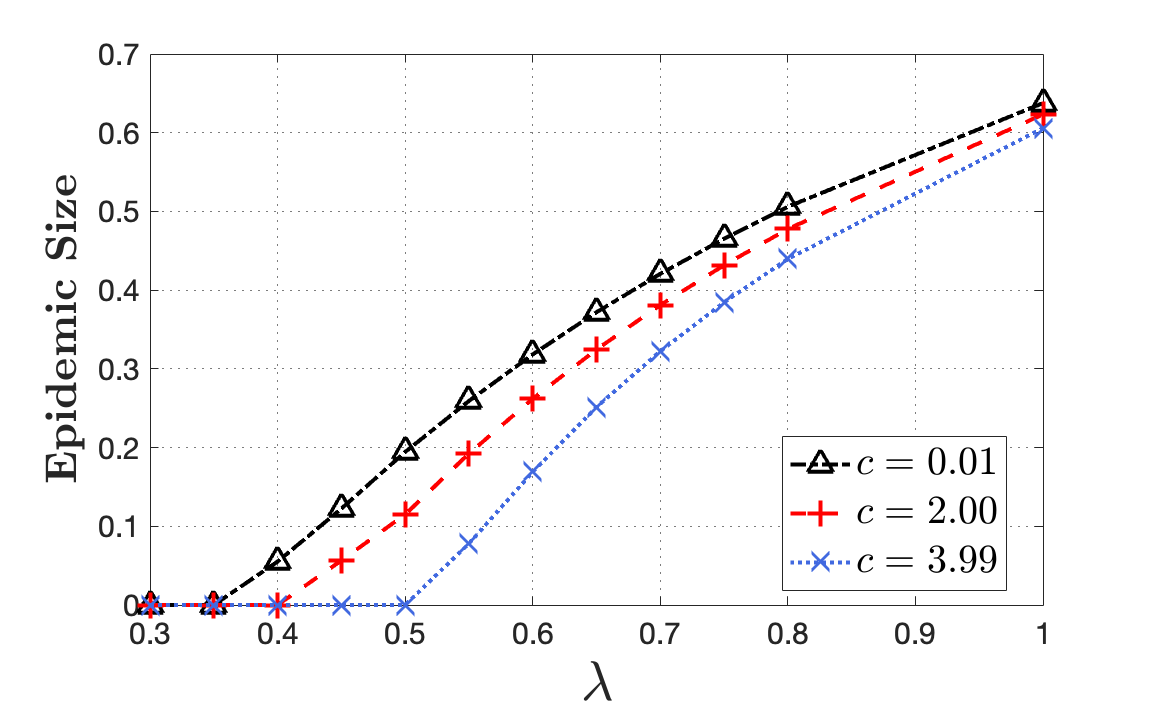}}
	\caption{\sl The impact of clustering: For each node, the number of incident single-edges is $2\times \mathrm{ Poi}\left(\frac{4-c}{2} \lambda \right)$ and the number of triangles is $\mathrm{Poi}\left( \frac{c}{2} \lambda \right)$. A higher value of $c$ indicates a larger clustering coefficient. The network size $n$ is $2 \times 10^5$, and the number of independent experiments for each data point is $10^4$. We set $T_1=0.2, T_2=0.5, \mu_{11}=\mu_{22}=0.75$. Our experimental results show that high clustering increases the threshold of epidemics and reduces the probability of emergence around the transition point.}\label{fig:clusterlambda}\end{figure*}

		Next, to evaluate the impact of clustering, we consider a joint degree distribution that allows us to control the level of clustering, while keeping the mean total degree fixed \cite{yong_tnse}. In particular, for each node, we set the number of incident single-edges as $2 \times \rm{ Poisson}\left(\frac{4-c}{2} \lambda \right)$ and the number of triangles to $\mathrm{Poisson}\left( \frac{c}{2} \lambda \right)$ where $c \in \left[0,4\right]$. This ensures that as $c$ varies, both the mean and the variance of the degree distribution remains constant, allowing us to isolate the effects of clustering. The parameter $c$ controls the level of clustering in the contact network. As $c$ increases, the clustering coefficient of the network also increases. Observe that when $c=0$, there will be no triangles in the network, and the clustering coefficient will be close to zero. In contrast, when $c=4$, there will be no single-edges in the network, and the clustering coefficient will be close to one. As $c$ increases, the clustering coefficient of the network also increases. In Figure~\ref{fig:clusterlambda}, we consider three different values for the parameter $c$, namely, $c=0.01$, $c=2.00$, and $c=3.99$, respectively to illustrate the impact of the clustering coefficient on the probability of emergence and the epidemic threshold. Our results reveal that high clustering increases the threshold of epidemics and reduces the probability of emergence and mean epidemic size beyond the phase transition point. These observations are consistent with the single-strain spreading \cite{yong_tnse,miller2009percolation} on clustered networks. 
	
		\subsection{Joint Impact of Clustering and Evolution}
		\label{subsec:contrast}
		Next, we discuss the interplay of clustering and evolution on the probability of emergence of epidemic outbreaks. We consider the case where the fitness landscape consists of two strains. The process starts when the population is introduced to the first strain (strain-1) which is moderately transmissible and initially dominant in the population. In contrast, the other strain (strain-2) is highly transmissible and initially absent in the population but has the risk of emerging through mutations in strain-1. For $\mu_{22}=1$ and $\mu_{12} \in (0,1)$, we have:
		\begin{align}
			\pmb{\mu}= \left[   \begin{matrix}
				1-\mu_{12} &  \mu_{12}\\ 
				0 &  1
			\end{matrix} \right] ;~~ \pmb{T}=
			\left[   \begin{matrix}
				T_1 &  0\\ 
				0 &  T_2
			\end{matrix} \right],~~ T_1<T_2.\label{eq:onesteprev}
		\end{align} 
		The above mutation and transmission parameters \eqref{eq:onesteprev} correspond to the \emph{one-step irreversible} mutation scheme, which is used widely \cite{antia2003role, ne:alexday} to model scenarios where a simple change is required for the contagion to evolve to a highly transmissible variant. We first isolate the impact of clustering in altering the epidemic threshold in the following Lemma.
		\begin{lemma} {\sl For multi-strain spreading with one-step irreversible mutations \eqref{eq:onesteprev} on clustered networks with doubly Poisson distribution \eqref{eq:doublypoisson}, the epidemic threshold is given as:
				\begin{align}
					\rho(\pmb{J}) = \lambda_s T_2 \times \left(1+\left(\dfrac{2\lambda_t}{\lambda_s}(1-T_2^2+T_2)\right) \right). \label{eq:lemma-one-step}
			\end{align}}
			\label{lem:lemma}
		\end{lemma}
		Through \eqref{eq:lemma-one-step}, we can see that compared to a network with only single-edges (distributed as $\rm Poisson (\lambda_s)$),  the epidemic threshold increases as a multiplicative factor of $\left(1+\left(\frac{2\lambda_t}{\lambda_s}(1-T_2^2+T_2)\right)\right)$. Moreover, when $\lambda_t=0$, we can recover the epidemic threshold of  $\rho(\pmb{J})=\lambda_s T_2$  corresponding to one-step irreversible mutations on a network with a vanishingly small clustering coefficient \cite{ne:alexday,ne:pnas}. A derivation of Lemma~\ref{lem:lemma} is presented in Section~\ref{subsec:lemma-derive}.

		\begin{figure}[t]
			\centering
			\includegraphics[scale=0.6]{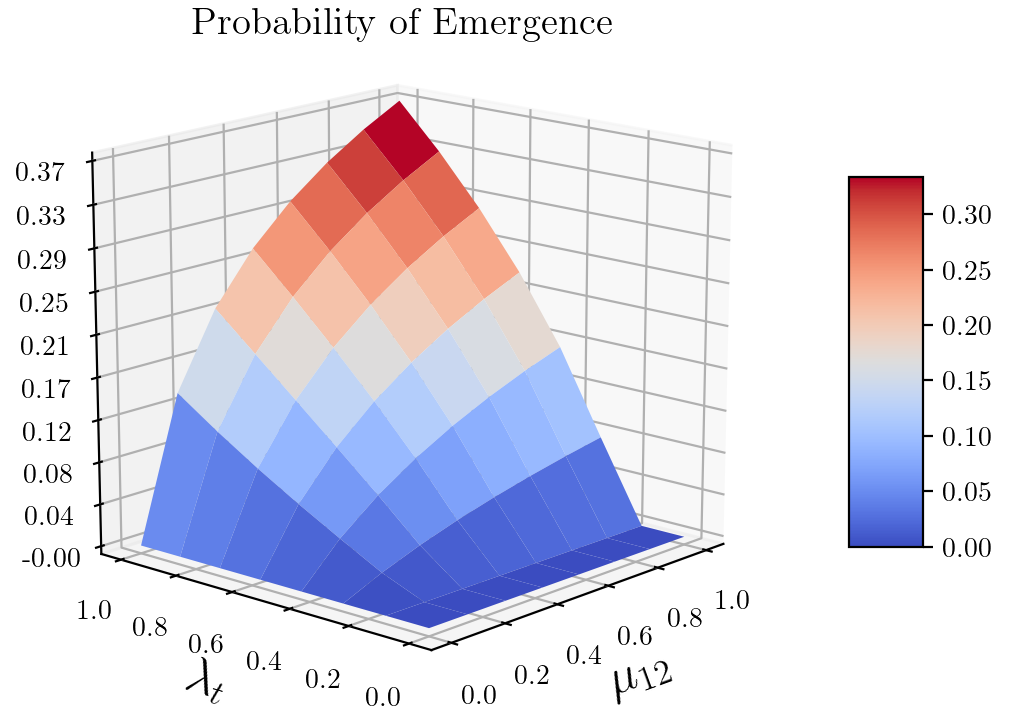}
			\caption{\sl  We consider a contact network following a doubly Poisson distribution \eqref{eq:doublypoisson} and with one-step irreversible mutations \eqref{eq:onesteprev}. we plot the probability of emergence given by Theorem~\ref{thm:prob} as a function of $(\lambda_t, \mu_{12})$. Here, $\lambda_t=0$ and $\mu_{12}=0$ respectively correspond to the absence of clustering and mutations. We set $T_1=0.2, T_2=0.7,\mu_{22} = 1, \lambda_s=1$. We observe that the joint impact of clustering and evolution $(\lambda_t>0,\mu_{12}>0)$ can lead to a significant increase in the probability of emergence of an epidemic as compared to the case when either of these phenomena act in isolation ($\lambda_t=0$ or $\mu_{12}=0$).}
			\vspace{-4mm}
			\label{fig:surf-prob}
		\end{figure} 
		In what follows, we note that the addition of clustering in the network structure can offer additional pathways for mutations which can, in turn, alter the course of the pandemic. We demonstrate this phenomenon for the case when the distribution of the single-edges and triangles is doubly Poisson with parameters $\lambda_s$ and $\lambda_t$ respectively, as in \eqref{eq:doublypoisson}. We vary $\lambda_t$ and  $\mu_{12}$ in the interval $[0,1]$ to parameterize the transition from a single-strain to a multi-strain spreading and from an unclustered to a clustered network. Note that when $\lambda_t = 0$, the network only comprises of single-edges and has a vanishingly small clustering coefficient \cite{ne:newman2018networks}. On the other hand, since the spreading process is initiated with strain-1, setting $\mu_{12} = 0$ corresponds to a single-strain setting.
		
		In Figure~\ref{fig:surf-prob}, we invoke Theorem~\ref{thm:prob} and plot the probability of emergence as a function of $(\lambda_t, \mu_{12})$, while setting $T_1=0.2, T_2=0.7,\mu_{22} = 1$, and $\lambda_s=1$.   We observe that the probability of emergence remains low when either $\mu_{12}=0$ \emph{or} $\lambda_t=0$, i.e., the likelihood of seeing an epidemic remains small when clustering or mutations act in isolation. In other words, even when mutations occur with a positive probability, the absence of triangles $(\lambda_t=0)$ renders a negligible risk of an epidemic outbreak. Similarly, in the absence of mutations $(\mu_{12}=0)$, clustering alone does not lead to an increased risk of an epidemic. However, for $\mu_{12}=\lambda_t=1$, i.e., when mutations occur with a high probability \emph{and} clustering is significant, we observe the probability of emergence rises sharply (Figure~\ref{fig:surf-prob}). This observation further highlights the need for evaluating risks of emergence of highly contagious mutations in the light of the structure of the contact network.
		
		\section{Proof Sketch}
		\label{sec:proof-sketch}
		\subsection{Outline for Proof of Theorem~\ref{thm:prob}}
		\label{subsec:thm-proof}
		To see intuitively why Theorem~\ref{thm:prob} holds, we note that $h_i(x)$ (respectively, $g_i(x)$) corresponds to the probability generating function (PGF) of the number of {\em finite} nodes reached and infected by following a randomly selected single-edge (respectively, triangle) emanating from a node carrying strain-$i$. This gives a way to define the PGF of the number of finite nodes reached and infected by selecting a node uniformly at random and making it type-$i$, denoted by $Q_i(x)$. Observe that  
		\begin{align}
			Q_i(x)  = x \sum_{s,t} q_{s,t} h_i(x)^s g_i(t)^t.
			\label{clustEvo_eq:rootEquation}
		\end{align}
		The validity of (\ref{clustEvo_eq:rootEquation}) can be seen as follows-- the factor $x$ accounts for the node which is selected randomly and given the infection as the seed of the process. Note that this node has a joint degree $(s,t)$ with probability $q_{s,t}$. Since this node carries strain-$i$, the number of nodes reached and infected by each of its $s$ single-edges (respectively, each of the $t$ triangles) has a generating function $h_i(x)$ (respectively, $g_i(x)$). From the {\em powers} property of PGFs \cite{newman2001random}, the total number of nodes reached and infected in this process when the initial node carries strain-$i$ and has joint degree $(s,t)$ has a generating function $h_i(x)^s g_i(x)^t$. As we average over all possible joint degrees $(s,t)$, we obtain (\ref{clustEvo_eq:rootEquation}). Note that when $Q(i)=1$, the number of infected nodes is {\em finite}  from the conservation of probability. Whereas when $Q(i)<1$, the corresponding probability $1-Q(i)$ gives the probability of infecting an {\em infinite} number of nodes leading to an epidemic outbreak. 
		
		 Next, we observe that for a node reached by following a single-edge (resp., triangle) selected uniformly at random, the joint degree distribution is proportional to the number of single-edges (resp., triangle) assigned to that node and given by $s q_{s,t} / \langle s \rangle$ (resp., $t q_{s,t} / \langle t \rangle$) where $\langle s \rangle=\sum_{s,t} s q_{s,t}$ (resp., $\langle t \rangle = \sum_{s,t} t q_{s,t}$) ensures normalization. We first state and explain how to derive $h_1(x)$, the PGF of the number of finite nodes reached and infected by following a randomly selected single-edge emanating from a node carrying strain-$1$;
		 	\begin{align}
h_1(x) &= 1-T_1+T_1 x \Big( \mu_{11} \sum_{s,t} \frac{s q_{s,t}}{\langle s \rangle} h_1(x)^{s-1} g_1(x)^t  \nonumber\\&\quad+  \mu_{12} \sum_{s,t} \frac{s q_{s,t}}{\langle s \rangle} h_2(x)^{s-1} g_2(x)^t \Big).
\label{clustEvo_eq:hEquation}
\end{align}
		 We note that if no transmission occurs along the randomly selected single-edge emanating from a node carrying strain-$1$, we get the factor of $(1-T_1)x^0$ in \eqref{clustEvo_eq:hEquation}. Whereas, if transmission occurs along the selected single-edge, the number of nodes eventually infected will be one {\em plus} all the nodes reached and infected due to node at the endpoint of the selected edge, which contributes a factor $T_1 x \Big( \mu_{11} \sum_{s,t} \frac{s q_{s,t}}{\langle s \rangle} h_1(x)^{s-1} g_1(x)^t+  \mu_{12} \sum_{s,t} \frac{s q_{s,t}}{\langle s \rangle} h_2(x)^{s-1} g_2(x)^t $\Big). This follows from noting that the node reached by following the randomly selected single-edge has already utilized one of its single-edges to connect to its parent, and it has $s-1$ remaining single-edges and $t$ triangles. Moreover, this node acquires strain-1 (resp., strain-2) with probability $\mu_{11}$ (resp., $\mu_{12}$). When this node carries strain-$1$ (resp., strain-$2$), the powers property of PGFs readily implies that the number of nodes infected due to this node has a generating function $h_1(x)^{s-1} g_1(x)^t$ (resp., $h_2(x)^{s-1} g_2(x)^t$).  Averaging over all possible joint degrees and node types yields \eqref{clustEvo_eq:hEquation}. Due to space constraints, we direct the reader to Section~\ref{sec:proof-details} for the derivation for the PGFs $h_1(x)$, $h_2(x)$, $g_1(x)$, and $g_2(x)$. 
		\subsection{Outline for Proof of Theorem~\ref{thm:threshold}}
		\label{subsec:proof-thresh}
		The proof of Theorem~\ref{thm:threshold} is linked to the stability of the fixed point solutions of (\ref{eq:ne:rec1}, \ref{eq:ne:rec2}). We note that for $i=1,2$, the set of equations (\ref{eq:ne:rec1}, \ref{eq:ne:rec2}) admits a trivial fixed point $h_1(1)=h_2(1)=g_1(1)=g_2(1)=1$. Substituting back into (\ref{clustEvo_eq:rootEquation}) gives $1-Q_i(1)=0$, i.e., all infected components are of finite size, and no outbreak emerges. To check the stability of this trivial solution, we linearize the set of equations (\ref{eq:ne:rec1}, \ref{eq:ne:rec2}) around $h_1(1)=h_2(1)=g_1(1)=g_2(1)=1$ and compute the corresponding Jacobian matrix $\pmb{J}=[J_{ij}]$. 
  We note that when the mutation matrix is {indecomposable}, i.e., each type of strain can eventually lead to the emergence of any other type with a positive probability, then the threshold theorem \cite{ne:alexday} guarantees extinction $h_1(1)=h_2(1)=g_1(1)=g_2(1)=1$ \cite{ne:alexday} if $\rho(\boldsymbol{J})\leq 1$.  Moreover, if $\rho(\pmb{J})>1$, then there exists another stable solution with $h_1(1), h_2(1), g_1(1), g_2(1)<1$, leading to a positive probability of emergence, i.e., $1-Q_i(1)>0$. Therefore, $\rho(\pmb{J})=1$ emerges as the epidemic threshold. For the {decomposable} case, the threshold theorem \cite{ne:alexday} guarantees extinction if $\rho(\boldsymbol{J})\leq 1$; however, the uniqueness of the fixed-point solution does not necessarily hold when $\rho(\boldsymbol{J})>1$.

  
				\section{Additional Proof Details}
    \label{sec:proof-details}
				\subsection{Deriving Probabilities in Table~\ref{table:tab2}}
				Here, we define the probabilities for the occurrence of the different configurations described in Table~\ref{table:tab2} and Figure~\ref{clustEvo_fig:fig1}, based on the endpoints of the triangle emanating from a parent node that carries strain-$i$. Throughout, we refer to a node carrying strain-$i$ as being a type-$i$ node.
				
				\begin{enumerate}
		{\item  $ p_{i1}$ \emph{(Neither end node was infected)}:  This configuration occurs when the parent node fails to infect both end nodes, which happens with probability $\left( 1- T_i \right)^2$.}
		
		{\item  $ p_{i2}$ \emph{(One end node was infected and has become type-$1$)}: This occurs when the parent node infects one of the end nodes, which later becomes type-$1$, {\em and} neither the parent node nor the infected end node succeeds in infecting the other end node.  Hence we have $2 T_i \mu_{i1} \left(1-T_i \right) \left(1-T_1 \right)$ as the associated probability, where the multiplication by $2$ is due to symmetry, i.e., either of the two end nodes could be the infected node.}
	
		{\item  $ p_{i3}$ \emph{(Both end nodes were infected and have become type-$1$)}: Note that we already derived this in Section~\ref{subsec:res-prelim}.}
		
		\item  $ p_{i4}$ \emph{(Only one end node was infected and has become type-$2$)}: This can be derived similar to $p_{i2}$, the probability of this configuration is $2 T_i \mu_{i2} \left(1-T_i \right) \left(1-T_2 \right)$.
		
		\item  $ p_{i5}$ \emph{(Both end nodes were infected and have become type-$2$)}: Akin to the derivation of $p_{i3}$, the probability of this configuration is $\left(T_i \mu_{i2} \right)^2 + 2T_i\mu_{i2}\left( 1- T_i \right) T_2 \mu_{22}$.
		
		\item  $ p_{i6}$ \emph{(Both end nodes were infected, one of them has become type-$1$, and the other has become type-$2$)}:This configuration occurs when the parent node infects both end nodes, then one of them becomes type-$1$ and the other becomes type-$2$, {\em or} the parent node infects {\em only} one node that later becomes type-$1$ (or type-$2$) and infects the other. Hence, the probability of this configuration is given by $2 \left( T_i^2 \mu_{i1} \mu_{i2} + T_i \mu_{i1} \left(1-T_i \right)T_1 \mu_{12}\right)$ $+2 \left(  T_i \mu_{i2} \left(1-T_i \right)T_2 \mu_{21}  \right)$
where the multiplication by the factor $2$ is again due to symmetry.	\end{enumerate}
We present detailed derivations for the relevant PGFs for establishing Theorem~\ref{thm:prob} in \ref{subsec:pgfdetail}.

\subsection{Proof of Lemma~\ref{lem:lemma}}
\label{subsec:lemma-derive}
We denote, $\beta_s=\frac{\langle s^2 \rangle - \langle s \rangle}{\langle s \rangle} $,  $\beta_t= \frac{\langle t^2 \rangle - \langle t \rangle}{\langle t \rangle}$, $\lambda_s =\langle s \rangle$ and $\lambda_t= \langle t \rangle$. When the distribution of single-edges and triangles are independent Poisson distributions\eqref{eq:doublypoisson}, we have $\beta_s=\lambda_s, \beta_t=\lambda_t$ and substituting into  \eqref{eq:JacobianMat}, we get
				\begin{align}
					\pmb{J}&= 	\left[
					\begin{matrix}
						\pmb{\Pi}&\pmb{0} \\
						\pmb{0} & \pmb{\Delta}
					\end{matrix}
					\right]
					\left[
					\begin{matrix}
						\lambda_s 	\pmb{\rm I}&\lambda_t \pmb{\rm I}\\
					\lambda_s  \pmb{\rm I}& \lambda_t \pmb{\rm I}
					\end{matrix}
					\right],\nonumber
				\end{align}
and further using \eqref{eq:onesteprev} and substituting $\mu_{22}=1, \mu_{21}=0$, have $\pmb{J}=$
\begin{equation}
	\left[
	\begin{matrix}
		T_1 (1-\mu_{12}) \lambda_s & T_1 \mu_{12} \lambda_s & T_1 (1-\mu_{12}) \lambda_t & T_1 \mu_{12} \lambda_t \\
		0& T_2 \lambda_s & 0&T_2  \lambda_t \\
		\Delta_{1,1} \lambda_s & \Delta_{1,2}  \lambda_s & \Delta_{1,1}\lambda_t & \Delta_{1,2}  \lambda_t \\
		0& \Delta_{2,2} \lambda_s & 0 & \Delta_{2,2}  \lambda_t 
	\end{matrix}
	\right]
	\label{eq:one-step-irrev-J-matrix}
	\nonumber
\end{equation}
The spectral radius of the above matrix is given as follows:
\begin{align}
	\rho(\pmb{J})&= \lambda_s\max\left\{T_1 (1-\mu_{12})+ \dfrac{\lambda_t}{\lambda_s}\Delta_{1,1} , T_2+ \dfrac{\lambda_t}{\lambda_s}\Delta_{2,2}  \right\}\nonumber\\
	&=  \lambda_s \left(T_2+ \dfrac{\lambda_t}{\lambda_s}\Delta_{2,2}  \right)\label{eq:conseq-of-t1t2},
\end{align}
where \eqref{eq:conseq-of-t1t2} is a consequence of the larger transmissibility of strain-$2$ \eqref{eq:onesteprev} which also implies $\Delta_{1,1}<\Delta_{2,2}$. Lastly, substituting $\mu_{22}=1$ in the definition of $\Delta_{22}$, we have
\begin{align}
   \Delta_{22}&= p_{24}+ 2p_{25}+ p_{26}\nonumber\\
   &=p_{24}+ 2p_{25} ~~~~~(\mu_{21}=0)\nonumber\\
   &=  2T_2  \left(1-T_2^2+T_2\right) \label{eq:conseq-of-t1-t2++}
\end{align}
Substituting~\eqref{eq:conseq-of-t1-t2++} in~\eqref{eq:conseq-of-t1t2}, we obtain~\eqref{eq:lemma-one-step}. 
		\section{Conclusions}
		\label{sec:conc}
We analyzed multi-strain spreading on networks with tuneable clustering and arbitrary degree distributions. We derived the probability of emergence of an epidemic outbreak and the critical {epidemic threshold} beyond which epidemics occur with a positive probability. We also present a heuristic solution that reduces the multi-strain spreading to single-strain spreading to estimate the conditional mean of the fraction of infected individuals given that an epidemic outbreak has occurred. Clustering within networks significantly impacts various epidemic characteristics. For contact networks characterized by a doubly Poisson distribution, we find that increasing the clustering coefficient while keeping the mean degree fixed raises the epidemic threshold, decreases the probability of emergence, and reduces the mean epidemic size. Furthermore, an analytical case study of one-step irreversible mutation patterns reveals that clustering can create additional pathways for mutations, thereby influencing the progression of an epidemic.
Our framework allows for evaluating the emergence of highly transmissible variants in relation to the extent of clustering in the network. Important future directions include leveraging the clustered multi-strain model on multi-layer networks, particularly those with higher-order community structures. Another key area of focus is investigating scenarios of co-infection and cases where exposure to one strain does not confer full immunity to another strain.

		\section{Acknowledgements}
This research was supported in part by the National Science Foundation through grant CCF-\#2225513 and by the Army Research Office through grant \#W911NF-22-1-0181. O.Y. also acknowledges the IBM academic award. M.S. acknowledges seed grants by the Cylab Security and Privacy Institute and the Center for Informed Democracy \& Social - cybersecurity (IDeaS), the Dowd Fellowship, Knight Fellowship, Lee-Stanziale Ohana Endowed Fellowship, and Cylab Presidential Fellowship at Carnegie Mellon University.


		\bibliographystyle{IEEEtran}
		\bibliography{ref}
  \section*{Author Biographies}
  \vspace{-40pt}
  \begin{IEEEbiography}[{\includegraphics[width=1.1in,height=1.05in,clip,keepaspectratio]{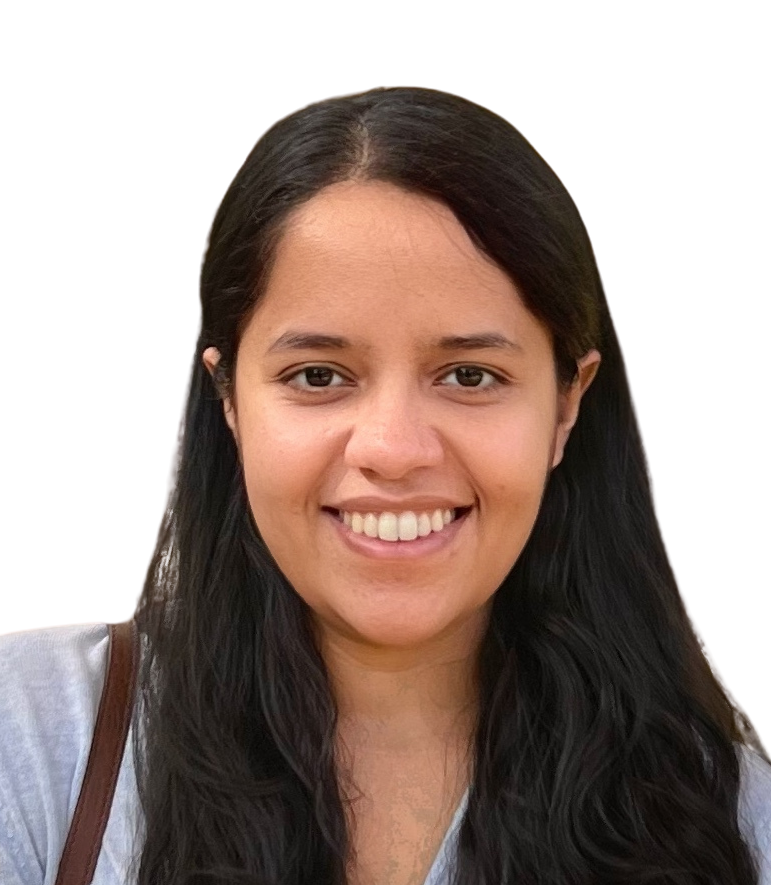}}]{Mansi Sood}
    (S'20) received the joint B.Tech. and M. Tech degree in Electrical Engineering from the Indian Institute of Technology (IIT) Bombay, India. She is currently a Ph.D. candidate in Electrical and Computer Engineering at Carnegie Mellon University (CMU). Her research interests span stochastic modeling, performance analysis, learning, and optimization. Her work won a Best Paper Award at the IEEE International Conference on Communications (ICC) '21. She has been recognized as a Schmidt Science Fellow '24 and twice recognized as an EECS Rising Star (MIT~'21 and Georgia Tech~'23).
\end{IEEEbiography}
  \begin{IEEEbiography}[{\includegraphics[width=0.8in,height=1.2in,clip,keepaspectratio]{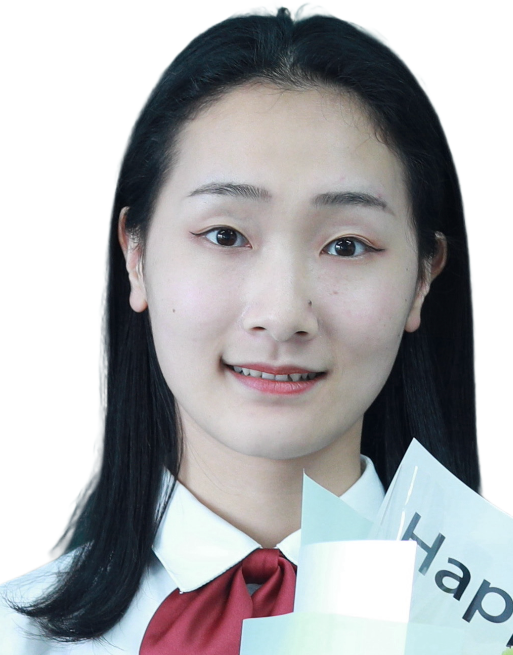}}]{Hejin Gu}
  received her master's degree in Software Engineering from Carnegie Mellon University. She also earned a BE in Computer Science from Zhejiang University.
\end{IEEEbiography}
  \begin{IEEEbiography}[{\includegraphics[width=1in,height=1.05in,clip,keepaspectratio]{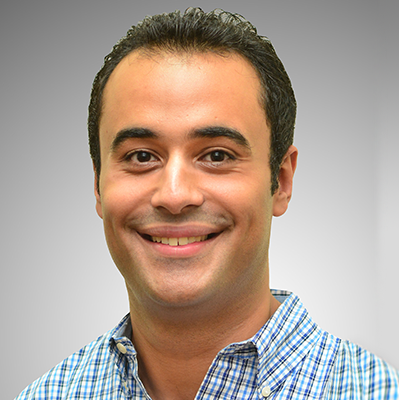}}]{Rashad Eletreby}
  holds a Ph.D. in Electrical and Computer Engineering from Carnegie Mellon University, where his research focused on random graph theory, network science, and machine learning. Currently, he serves as a Senior Manager of Data Science at Walmart Global Tech, leading a team that explores state-of-the-art machine learning algorithms to support Walmart's accelerated marketplace initiatives. Prior to his role at Walmart, Dr. Eletreby was a Principal Data Scientist at Rocket Travel by Agoda, where he contributed to advancements in neural translation, ranking systems, and A/B testing methodologies.
\end{IEEEbiography}
  \begin{IEEEbiography}[{\includegraphics[width=1in,height=1.05in,clip,keepaspectratio]{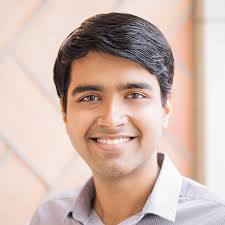}}]{Swarun Kumar}
   is the Sathaye Family Foundation Professor at Carnegie Mellon
University's ECE department, with a courtesy appointment in the CS department, the HCI institute and Cylab.
His research builds next-generation wireless network protocols and services. He leads the Emerging Wireless
Technologies (WiTech) laboratory at CMU. He is a recipient of the 2024 ACM SIGMOBILE Rockstar award, the
2021 ACM SIGBED Early Career Researcher Award, the NSF CAREER Award and the Google Faculty Research
award. He has also received the George Sprowls Award for best Ph.D. thesis in Computer Science at MIT and
the President of India gold medal at IIT Madras.
\end{IEEEbiography}
  \begin{IEEEbiography}[{\includegraphics[width=1in,height=1.05in,clip,keepaspectratio]{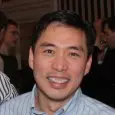}}]{Chai Wah Wu}
  (S’88-M’96-SM’00-F’01) received his B.S. degree in Computer Engineering and B.A. degree in Cognitive Science 
from Lehigh University and M.A. degree in Mathematics and M.S/Ph.D degrees in Electrical Engineering from 
the University of California at Berkeley.  He has been with IBM T. J. Watson Research 
Center since 1996 and is currently a Principal Research Scientist.
His research interests include nonlinear dynamics of networked circuits and systems,
 algebraic graph theory, approximate computing, machine learning, and multimedia security.  
He has published over 200 papers and was granted 78 U.S. patents and 
has served as chair and technical program committee member for several international 
conferences.  He was an associate editor of IEEE TCAS, Editor-in-Chief of IEEE CASM and served on the CAS Board of Governors.  
He has been serving as an IEEE EAB Program Evaluator since 2006, is an INFORMS Certified Analytics Professional and was elected Fellow of the IEEE in 2001. 
\end{IEEEbiography}
  \begin{IEEEbiography}[{\includegraphics[width=1in,height=1.2in,clip,keepaspectratio]{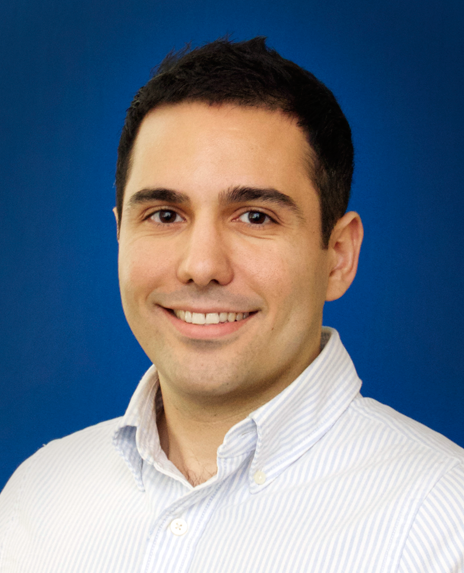}}]{Osman Ya\u{g}an}
  is  a Full Research Professor of Electrical and Computer Engineering (ECE) at Carnegie Mellon University (CMU), where he is also an affiliate faculty in the School of Computer Science and a core member of CyLab Security and Privacy Institute. He received his Ph.D. degree in Electrical and Computer Engineering from the University of Maryland at College Park, MD in 2011, and his B.S. degree in Electrical and Electronics Engineering from the Middle East Technical University, Ankara (Turkey) in 2007.
His research focuses on modeling, analysis, and performance optimization of computing systems, and uses tools from applied probability, network science, data science, and machine learning. 

Dr. Yağan is a senior member of IEEE, and a recipient of a CIT Dean's Early Career Fellowship, an IBM Academic Award, and best paper awards in ICC 2021, IPSN 2022, and ASONAM 2023.

\end{IEEEbiography}
\section{Appendix}
\subsection{Deriving PGFs $h_i(x), g_i(x)$, $i=1,2$ in Theorem~\ref{thm:prob}}
\label{subsec:pgfdetail}
\subsubsection{Deriving $h_1(x)$ and $h_2(x)$}
We start by deriving an expression for $h_1(x)$. Note that $h_1(x)$ denotes the probability generating function of the number of finite nodes reached and infected by following a randomly selected single edge emanating from a type-$1$ node. Observe that if this edge is not occupied (an event which happens with probability $1-T_1$), then no node whatsoever would receive the infection following this edge (leading to a term $(1-T_1)x^0$ in the generating function for $h_1(x)$). If this edge is occupied (an event that happens with probability $T_1$), then the current node must have received an infection with strain-$1$, and it would either become type-$1$ if the pathogen does not mutate (an event that happens with probability $\mu_{11}$) or type-$2$ if the pathogen mutates to strain-$2$ (an event that happens with probability $\mu_{12}$). Averaging over all possible mutation outcomes, we obtain \eqref{clustEvo_eq:hEquation}.

The validity of (\ref{clustEvo_eq:hEquation}) can be seen as follows. When the node under consideration receives the infection (which happens when the edge is occupied), the number of nodes reached and infected will be one {\em plus} all the nodes reached and infected due to the particular node under consideration. This node could be type-$1$ with probability $\mu_{11}$ or type-$2$ with probability $\mu_{12}$. In either case, the probability that this node has a joint degree $(s,t)$ would be given by ${s q_{s,t}}/{\langle s \rangle}$ since it is already known that this node has at least one single-edge. Since this node has already utilized one of its single-edges to connect to its parent, it has $s-1$ remaining single-edges and $t$ triangles that it could utilize to spread the infection. When the node is type-$1$ (respectively, type-$2$), the powers property of generating functions readily implies that the number of nodes reached and infected due to this node has a generating function $h_1(x)^{s-1} g_1(x)^t$ (respectively, $h_2(x)^{s-1} g_2(x)^t$). Averaging over all possible joint degrees and node types gives (\ref{clustEvo_eq:hEquation}). Similarly, we derive an expression for $h_2(x)$ as follows.
\begin{align}
h_2(x) &= 1-T_2+T_2 x \Big( \mu_{21} \sum_{s,t} \frac{s q_{s,t}}{\langle s \rangle} h_1(x)^{s-1} g_1(x)^t  \nonumber\\&\qquad+  \mu_{22} \sum_{s,t} \frac{s q_{s,t}}{\langle s \rangle} h_2(x)^{s-1} g_2(x)^t \Big)
\label{clustEvo_eq:hEquation_part2}
\end{align}

\subsubsection{Deriving $g_1(x)$ and $g_2(x)$}
The situation becomes more challenging as we consider triangles since we need to jointly consider the status of the two end nodes of a triangle and we need to consider different cases, as seen in Figure~\ref{clustEvo_fig:fig1}. In reference to Table~\ref{table:tab2}, let $p_{ij}$ denotes the probability of the $j$th configuration when the parent node is type-$i$ for $i=1,2$ and $j=1,\ldots,6$. We then have
\begin{align}
g_i(x)  
&=p_{i1} + p_{i2}  \left( x \sum_{s,t} \frac{t q_{s,t}}{\langle t \rangle} h_1(x)^s g_1(x)^{t-1}\right)\nonumber\\&\quad+ p_{i3} \left( x \sum_{s,t} \frac{t q_{s,t}}{\langle t \rangle} h_1(x)^s g_1(x)^{t-1} \right)^2 \nonumber\\&\quad+  p_{i4}  \left( x \sum_{s,t} \frac{t q_{s,t}}{\langle t \rangle} h_2(x)^s g_2(x)^{t-1} \right) \nonumber\\&\quad+ p_{i5} \left( x \sum_{s,t} \frac{t q_{s,t}}{\langle t \rangle} h_2(x)^s g_2(x)^{t-1} \right)^2  \nonumber\\&\quad+  p_{i6} \left( x \sum_{s,t} \frac{t q_{s,t}}{\langle t \rangle} h_1(x)^s g_1(x)^{t-1} \right)\nonumber\\&\qquad\cdot \left( x \sum_{s,t} \frac{t q_{s,t}}{\langle t \rangle} h_2(x)^s g_2(x)^{t-1} \right), \label{clustEvo_eq:triangleEq}
\end{align}
for $i=1,2$, where the validity of (\ref{clustEvo_eq:triangleEq}) can be seen as follows. With probability, $p_{i1}$,  both end nodes in the triangle do not propagate the infection. This leads to a term $p_{i1} x^0$ in the PGF. Next, with probability $p_{i2}$ (respectively, $p_{i4}$), the triangle is in a configuration corresponding to one end node having acquired strain-1 (respectively, strain-$2$). In the above case, the degree distribution of the infected node is given by ${t q_{s,t}}/{\langle t \rangle}$ since it is already known that this node has at least one triangle. Since this node has already utilized one triangle to connect to its parent, it can only utilize the remaining $t-1$ triangles and $s$ single-edges to infect its neighbors. Using the powers property of PGFs, along with the fact that in this configuration, the node under consideration is type-$1$ (respectively, type-$2$), the PGF for the number of subsequent infections would be given by $h_1(x)^s g_1(x)^{t-1}$ (respectively, $h_2(x)^s g_2(x)^{t-1}$). 
For the configurations corresponding to probabilities $p_{i3}$, $p_{i5}$, and $p_{i6}$, the two end nodes spread the infection, albeit to two independent sets of other nodes, hence we can utilize the powers property of PGFs as above to obtain the corresponding terms to obtain \eqref{clustEvo_eq:triangleEq}.
\subsection{Additional details for proof of Theorem~\ref{thm:threshold}}
\label{subsec:additional-thm-thresh}
Here we discuss the steps to linearize the set of equations (\ref{eq:ne:rec1}, \ref{eq:ne:rec2}) around $h_1(1)=h_2(1)=g_1(1)=g_2(1)=1$ and arrive at the corresponding Jacobian matrix $\pmb{J}=[J_{ij}]$ given in Theorem~\ref{thm:threshold}. In what follows, we show the form of the Jacobian matrix $\pmb{J}$. For notational simplicity, we define
\begin{align}
&f_1(h_1,h_2,g_1,g_2):=\nonumber\\
&1-T_1+T_1 x \left( \mu_{11} \sum_{s,t} \frac{s q_{s,t}}{\langle s \rangle} h_1^{s-1} g_1^t  +  \mu_{12} \sum_{s,t} \frac{s q_{s,t}}{\langle s \rangle} h_2^{s-1} g_2^t \right) \nonumber \\
&f_2(h_1,h_2,g_1,g_2):=\nonumber\\
&1-T_2+T_2 x \left( \mu_{21} \sum_{s,t} \frac{s q_{s,t}}{\langle s \rangle} h_1^{s-1} g_1^t  +  \mu_{22} \sum_{s,t} \frac{s q_{s,t}}{\langle s \rangle} h_2^{s-1} g_2^t \right) \nonumber \\
&f_3(h_1,h_2,g_1,g_2):=\nonumber\\
&p_{11} + p_{12} x \sum_{s,t} \frac{t q_{s,t}}{\langle t \rangle} h_1^s g_1^{t-1} + p_{13} \left( x \sum_{s,t} \frac{t q_{s,t}}{\langle t \rangle} h_1^s g_1^{t-1} \right)^2 + \nonumber \\
& \quad p_{14} x \sum_{s,t} \frac{t q_{s,t}}{\langle t \rangle} h_2^s g_2^{t-1} + c_{15} \left( x \sum_{s,t} \frac{t q_{s,t}}{\langle t \rangle} h_2^s g_2^{t-1} \right)^2 + \nonumber \\
& \quad p_{16} \left( x \sum_{s,t} \frac{t q_{s,t}}{\langle t \rangle} h_1^s g_1^{t-1} \right) \left( x \sum_{s,t} \frac{t q_{s,t}}{\langle t \rangle} h_2^s g_2^{t-1} \right) \nonumber \\
&f_4(h_1,h_2,g_1,g_2):=\nonumber\\
&p_{21}+ p_{22}x \sum_{s,t} \frac{t q_{s,t}}{\langle t \rangle} h_1^s g_1^{t-1} + p_{23}\left( x \sum_{s,t} \frac{t q_{s,t}}{\langle t \rangle} h_1^s g_1^{t-1} \right)^2 + \nonumber \\
& \quad p_{24}x \sum_{s,t} \frac{t q_{s,t}}{\langle t \rangle} h_2^s g_2^{t-1} + p_{25}\left( x \sum_{s,t} \frac{t q_{s,t}}{\langle t \rangle} h_2^s g_2^{t-1} \right)^2 + \nonumber \\
& \quad p_{26}\left( x \sum_{s,t} \frac{t q_{s,t}}{\langle t \rangle} h_1^s g_1^{t-1} \right) \left( x \sum_{s,t} \frac{t q_{s,t}}{\langle t \rangle} h_2^s g_2^{t-1} \right) \nonumber 
\end{align}
Linearizing around $h_1=h_2=g_1=g_2=1$, we obtain
\begin{align}
J_{i1}&=\frac{\partial}{\partial h_1} f_1(h_1, h_2, g_1, g_2)|{h_1=h_2=g_1=g_2=1} \nonumber \\
J_{i2}&=\frac{\partial}{\partial h_2} f_2(h_1, h_2, g_1, g_2)|{h_1=h_2=g_1=g_2=1} \nonumber \\
J_{i3}&=\frac{\partial}{\partial g_1} f_3(h_1, h_2, g_1, g_2)|{h_1=h_2=g_1=g_2=1} \nonumber \\
J_{i4}&=\frac{\partial}{\partial g_2} f_4(h_1, h_2, g_1, g_2)|{h_1=h_2=g_1=g_2=1} \nonumber 
\end{align}
for $i=1,2,3,4$. Additionally with \eqref{eq:pianddelta}, it follows that
\begin{align}
\pmb{J}&=
\left[
\begin{matrix}
 \Pi_{11}  \frac{\langle s^2 \rangle - \langle s \rangle}{\langle s \rangle}&  \Pi_{12}  \frac{\langle s^2 \rangle - \langle s \rangle}{\langle s \rangle} &\Pi_{11} \frac{\langle st \rangle}{\langle s \rangle} & \Pi_{12} \frac{\langle st \rangle}{\langle s \rangle} \\
\Pi_{21} \frac{\langle s^2 \rangle - \langle s \rangle}{\langle s \rangle} &\Pi_{22} \frac{\langle s^2 \rangle - \langle s \rangle}{\langle s \rangle} & \Pi_{21} \frac{\langle st \rangle}{\langle s \rangle} &\Pi_{22} \frac{\langle st \rangle}{\langle s \rangle} \\
	\Delta_{11} \frac{\langle st \rangle}{\langle t \rangle} & \Delta_{12} \frac{\langle st \rangle}{\langle t \rangle} & 	\Delta_{11} \frac{\langle t^2 \rangle - \langle t \rangle}{\langle t \rangle} &  \Delta_{12} \frac{\langle t^2 \rangle - \langle t \rangle}{\langle t \rangle} \\
\Delta_{21} \frac{\langle st \rangle}{\langle t \rangle} &  \Delta_{22}\frac{\langle st \rangle}{\langle t \rangle} & 	\Delta_{21}\frac{\langle t^2 \rangle - \langle t \rangle}{\langle t \rangle} &  \Delta_{22} \frac{\langle t^2 \rangle - \langle t \rangle}{\langle t \rangle}
\end{matrix} 
\right]\nonumber\\&= 	\left[
					\begin{matrix}
						\pmb{\Pi}&\pmb{0} \\
						\pmb{0} & \pmb{\Delta}
					\end{matrix}
					\right]
					\left[
					\begin{matrix}
						\frac{\langle s^2 \rangle - \langle s \rangle}{\langle s \rangle} 	\pmb{\rm I}&\frac{\langle st \rangle}{\langle s \rangle}  \pmb{\rm I}\\
						\frac{\langle st \rangle}{\langle t \rangle}   \pmb{\rm I}& \frac{\langle t^2 \rangle - \langle t \rangle}{\langle t \rangle} \pmb{\rm I}
					\end{matrix}
					\right].
\nonumber
\end{align}

\end{document}